\documentclass[preprint,12pt]{elsarticle}

\usepackage{amssymb}
\usepackage{todonotes}
\setuptodonotes{inline}
\usepackage{graphicx}
\usepackage{booktabs}
\usepackage{tabularx}
\usepackage{hyperref}
\usepackage{paralist}
\usepackage{tabulary}
\usepackage{ragged2e}
\usepackage{wrapfig}
\usepackage{comment}
\usepackage{listings}
\usepackage{caption}
\usepackage{subcaption}
\usepackage{multirow}
\usepackage{xurl}
\usepackage{marginnote}
\usepackage{float}

\newcolumntype{L}[1]{>{\raggedright\let\newline\\\arraybackslash\hspace{0pt}}m{#1}}
\newcolumntype{R}[1]{>{\raggedleft\let\newline\\\arraybackslash\hspace{0pt}}m{#1}}

\newcommand{\update}[3][0em]{#3}
\newcommand{\secondupdate}[3][0em]{#3}

\journal{Information Systems}

\begin{document}

\begin{frontmatter}



\title{\update{R1.2}{A Task Taxonomy for Conformance Checking}}


\author[inst1]{Jana-Rebecca Rehse}

\author[inst1]{Michael Grohs}

\author[inst2,inst3]{Finn Klessascheck}

\author[inst1]{Lisa-Marie Klein}

\author[inst4]{Tatiana von Landesberger}

\author[inst2,inst3]{Luise Pufahl}

\affiliation[inst1]{organization={University of Mannheim},
            addressline={L15, 1-6}, 
            postcode={68161}, 
            city={Mannheim},
            state={BW},
            country={Germany}}

\affiliation[inst2]{organization={Technical University of Munich, School of CIT},
            addressline={Bildungscampus 2}, 
            city={Heilbronn},
            postcode={74076}, 
            state={BW},
            country={Germany}}

\affiliation[inst3]{organization={Weizenbaum Institute},
            addressline={Hardenbergstr. 32}, 
            city={Berlin},
            postcode={10623}, 
            state={Berlin},
            country={Germany}}

\affiliation[inst4]{organization={University of Cologne},
            addressline={Albertus-Magnus-Platz}, 
            city={Cologne},
            postcode={50923}, 
            state={NRW},
            country={Germany}}

\begin{abstract}
Conformance checking is a sub-discipline of process mining, which compares observed process traces with a process model to analyze whether the process execution conforms with or deviates from the process design. 
Organizations can leverage this analysis, for example to check whether their processes comply with internal or external regulations or to identify potential improvements. 
\secondupdate{R2}{
Gaining these insights requires suitable visualizations, which make complex results accessible and actionable. 
So far, however, the development of conformance checking visualizations has largely been left to tool vendors.
As a result, current tools offer a wide variety of visual representations for conformance checking, but the analytical purposes they serve often remain unclear.
However, without a systematic understanding of these purposes, it is difficult to evaluate the visualizations’ usefulness. 
Such an evaluation hence requires a deeper understanding of conformance checking as an analysis domain.
To this end, we propose a task taxonomy, which categorizes the tasks that can occur when conducting conformance checking analyses. 
}
This taxonomy supports researchers in determining the purpose of visualizations, specifying relevant conformance checking tasks in terms of their goal, means, constraint type, data characteristics, data target, and data cardinality.
Combining concepts from process mining and visual analytics, we address researchers from both disciplines to enable and support closer collaborations.

\end{abstract}

\begin{highlights}
\item Presents a visualization task taxonomy for the conformance checking domain
\item Based on the structured analysis of 33 process mining case studies, which describe 102 tasks
\item Categorizes tasks along six dimensions: goal, means, constraint type, data characteristics, data target, data cardinality
\item Analyzes the frequency of the individual tasks as well as the dependencies between them
\item \secondupdate{R.2}{Provides foundations for designing and evaluating new visualizations}
\end{highlights}


\end{frontmatter}



\section{Introduction}

Process mining refers to a collection of data analysis techniques aimed at gaining insights on business processes in organizations \cite{aalst_2022_overview}. To this end, process mining techniques analyze so-called process event logs, which capture the executions of business processes and can be extracted from enterprise information systems. 
A well-established sub-discipline of process mining is conformance checking, which focuses on the relation between the prescribed process behavior and the process behavior that actually occurs in reality \cite{carmona2018conformance}. 
Conformance checking techniques compare the prescribed (to-be) process, typically specified in terms of a process model or a set of rules, with the actual process executions that are captured in the event log. 
This way, they are able to analyze whether a process execution conforms with or deviates from the prescribed process and where potential deviations occur~\cite{dunzer_2019_conformance_checking}. 
Organizations can leverage this analysis, for example to check whether their processes comply with internal or external regulations or to identify improvement potentials in their processes. 
Those analytical capabilities make conformance checking relevant for organizations in many different domains~\cite{emamjome_2019_case_study}. 
A recent study among practitioners even found that conformance checking is a critical feature in the selection of a process mining tool~\cite{survey_2021}. 

Despite these potentials and its alleged relevance, however, conformance checking is still not widely applied in practice \cite[p.~39]{reinkemeyer2020process}.
This lack of adoption can partially be explained by the high computational efforts of many conformance checking techniques, their dependency on process models, which may not always be available, and their focus on control-flow aspects, which is only one of the perspectives that are relevant for practice~\cite{voglhofer2020collection}. 
Over the last few years, researchers have tried to address these problems. 
For example, the computational efficiency of conformance checking techniques has considerably improved, to allow for applying them to large, real-life event logs \cite{dunzer_2019_conformance_checking}.
To reduce the dependency on available process models, researchers have looked into alternative approaches, e.g., by means of machine learning \cite{peeperkorn2023global}.
Conformance checking has also been extended beyond a pure control flow view to consider prescribed process behavior with regard to, e.g., the available resources and the required data~\cite{knuplesch_2017_framework,gall_2017_startISCViz,voglhofer2020collection}.


\secondupdate{R2}{
In this paper, we focus on another aspect that could support the adoption of conformance checking in practice~\cite{garcia_2017_complete,gschwandtner_2017_visual,klinkmuller_2019_mining}: the visualizations used to communicate results. 
For intricate data analyses like conformance checking, visualizations are essential for making results accessible and actionable because they enable users to quickly interpret findings and make informed decisions~\cite{Keim2008_visual}.
However, in process mining and particularly in conformance checking, the development of visualizations has so far been largely left to tool vendors. 
As a result, current process mining tools offer a wide variety of visual representations, which differ considerably in both the visual idioms employed and the type of information conveyed~\cite{rehse_2022_visualization,hage2024taxonomy}. 

Facing this variety of conformance checking visualizations, practitioners and tool vendors need guidance on making the right choices for their organizations.
Academic research can support them by evaluating the efficacy of visualizations and providing suggestions on their improvement. 
However, such an evaluation is often hard to conduct because researchers cannot establish a clear match between a given visualization and the specific analytical purpose it serves.
Although the importance of process mining visualizations has long been acknowledged~\cite{garcia_2017_complete,gschwandtner_2017_visual,klinkmuller_2019_mining}, systematic research on visualization design for conformance checking has only recently gained traction~\cite{rehse_2022_visualization,klessascheck2024StructuralModel}.
Hence, there is little empirical evidence to support decisions about which visualizations best fulfill particular user needs \cite{garcia_2017_complete,klessascheck2024StructuralModel}, making it difficult to provide informed guidance.

Against this backdrop, we argue that a thorough evaluation of existing conformance checking visualization requires a more thorough understanding of conformance checking as a process analysis domain.
Simply put, we need to know why and for what purposes process analysts use conformance checking.
Without a systematic understanding of these purposes, it is difficult to evaluate the usefulness or adequacy of the provided visualizations as we do not know what information a corresponding visualization should convey or which visualization idioms are best suited for this purposes.
Thus, evaluating visualizations and potentially developing new ones first requires a deeper understanding of conformance checking as an analysis domain \cite{munzner2009nested,sedlmair2012design}. 
}

To develop such a deeper domain understanding, we can draw on a well-established tool from visualization research: \emph{task taxonomies} \cite{kerracher2017constructing}. 
A \emph{task} refers to a specific question to be answered within the context of a data analysis domain. 
A task taxonomy systematically categorizes these tasks for this domain, providing a structured way to analyze user needs, which can then guide the design and evaluation of new visualization systems.
In the context of conformance checking, this is precisely what we are missing: a clear understanding of the tasks that users seek to perform with conformance checking. Without this understanding, it is difficult to design visualizations that are truly useful in practice \cite{klinkmuller_2019_mining}.

\update{R1.2, R2.2}{ 
Therefore, in this paper, we develop a task taxonomy for conformance checking. 
By identifying relevant conformance checking tasks and systematically categorizing them into multiple dimensions, this taxonomy provides a systematic understanding of the conformance checking domain by specifying and structuring the analysis questions that conformance checking can answer.}
To this end, this paper is structured as follows. 
In \autoref{sec:background}, we provide the necessary background on conformance checking. \autoref{sec:method} elaborates on the research method that we used to develop the task taxonomy, presented in \autoref{sec:taxonomy}. 
\update{R2.3}{\autoref{sec:illustrative} provides illustrative examples for identified tasks.} We discuss related work in \autoref{sec:related}, before concluding the paper in \autoref{sec:conclusion}.

\section{Background}
\label{sec:background}
\update{R2.4}{This section introduces the essential preliminaries for a task taxonomy for conformance checking. 
We explain the data sources for conformance checking followed by an introduction into conformance checking techniques.}

\begin{figure}[htb]
    \centering
    \includegraphics[width=.8\textwidth]{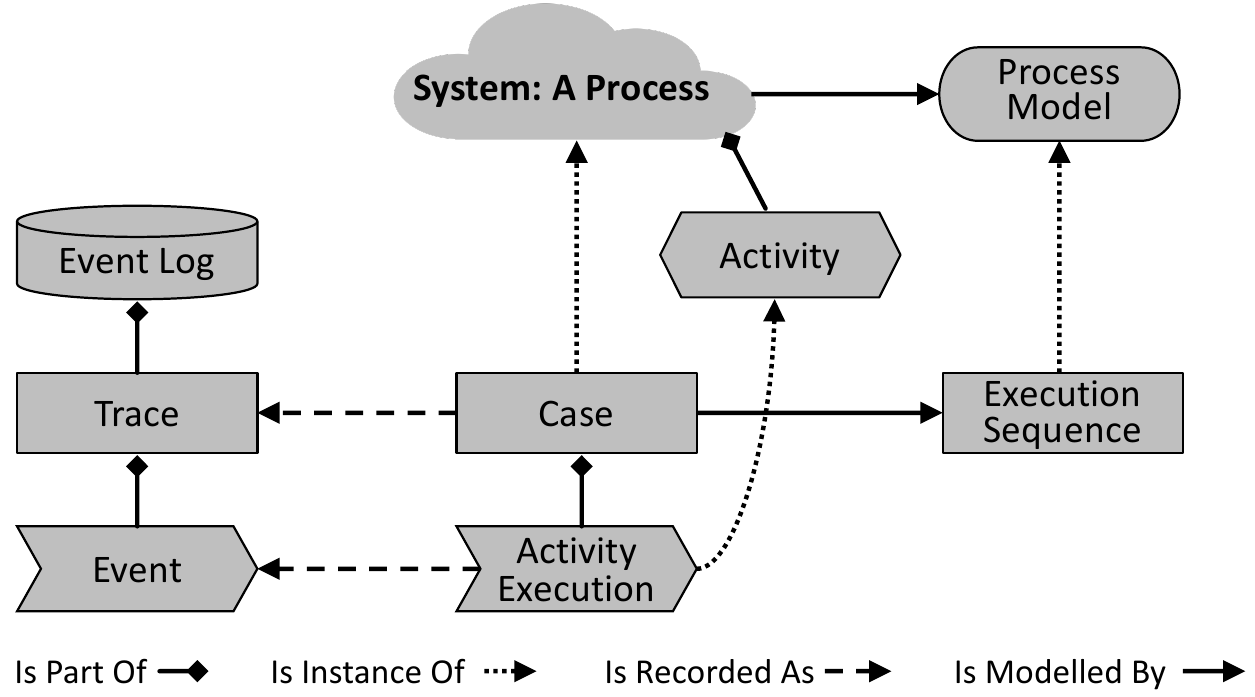}
    \caption{Overview of Business Process Terminology (figure adapted from \cite{carmona2018conformance})}
    \label{fig:terminology}
\end{figure}

\update{R2.4}{\subsection{Data Sources for Conformance Checking}
In the following, we elaborate on business processes and their models. Then, we introduce event logs that capture data on the real-world execution of a business process.  \autoref{fig:terminology} provides an overview over used terminology.}

\begin{figure}[htb]
    \centering
    \includegraphics[width=.8\textwidth]{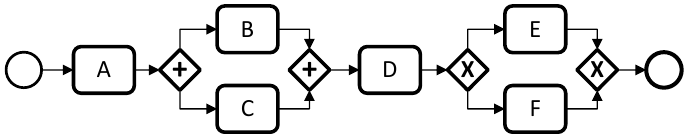}
    \caption{Exemplary Process Model in BPMN Notation}
    \label{fig:ex_model}
\end{figure}

\paragraph{Business Process \& Business Process Model} 
Organizations conduct a multitude of business processes to deliver products and services to customers, such as order-to-fulfill or purchase-to-pay. A \emph{business process}, embedded in an organizational and technical environment, consists of a set of activities executed to achieve a specific \emph{process outcome}~\cite{weske19}, such as fulfilling an order. Thereby, an \emph{activity} is a work unit that requires time to complete. Business processes are executed multiple times to handle different \emph{cases}, such as different orders. 
For handling the complexity of business processes, documenting, analyzing, redesigning them, and supporting their implementation in IT systems, a process can be captured in a \emph{process model} with the help of a modeling notation~\cite{dumas2018fundamentals}, such as BPMN (Business Process Model and Notation)~\cite{BPMN}.
As shown in \autoref{fig:terminology}, a process model serves as an abstract representation of the process for specific modeling goals and describes the allowed \emph{execution sequences} for different process cases~\cite{weske19}. 
As an example, consider the process model in \autoref{fig:ex_model} given as BPMN diagram. The process starts with activity A, after which both activities B and C are executed in any order, i.e., they are concurrent. Following the execution of activity D, the process is finished through either activity E or F.

A model can capture various process perspectives~\cite[Ch.~3]{dumas2018fundamentals}. The basis is the \emph{control-flow} perspective, which represents the order of activities (as shown in \autoref{fig:ex_model}). Additionally, the \emph{resource} perspective (responsible executors of activities), the \emph{data} perspective (documents and data used and produced by activities), and the \emph{time} perspective (time constraints for activities) can also be modeled. 

\paragraph{Event Log} During process execution, data on the different cases and their corresponding execution sequences is often recorded by IT systems in a so-called \emph{event log}.
As visualized in \autoref{fig:terminology}, an event log consists of a set of \emph{traces} and a set of \emph{events}. 
Events can represent different lifecycle changes of an activity, such as its enablement, start, or termination. An event includes information on which activity (lifecycle state) was executed, when it was executed, i.e., its timestamp, and to which case it belongs.
For example, the event \texttt{e1} given in Table~\ref{tab:ex_log} shows that activity \texttt{A} was started on the 1st of January 2024 at 9:00 am for the case with the identifier \texttt{id-4}. 
All events related to one process case form a \emph{trace}, e.g., \texttt{(A,B,C,D,F)} for the case \texttt{id-4}.
In the event log in Table~\ref{tab:ex_log}, there are nine events that belong to two traces (id-4 and id-7).
All traces with the same order of events can be summarized as a trace \emph{variant}.


\begin{table}[ht]
\centering
\begin{tabular}{llll}
\toprule
\textbf{Event} & \textbf{CaseID} & \textbf{Activity (start)} & \textbf{Timestamp} \\ \midrule
$e_1$ & id-4 & A & 01.01.24 09:00 \\
$e_2$ & id-4 & B & 01.01.24 09:14 \\
$e_3$ & id-4 & C & 01.01.24 09:28 \\
$e_4$ & id-4 & D & 01.01.24 09:43 \\
$e_5$ & id-4 & F & 01.01.24 09:57 \\
$e_6$ & id-7 & A & 01.01.24 11:12 \\
$e_7$ & id-7 & C & 01.01.24 11:55 \\
$e_8$ & id-7 & B & 01.01.24 12:38 \\
$e_9$ & id-7 & E & 01.01.24 13:21 \\ \bottomrule
\end{tabular}
\caption{Exemplary Event Log}\label{tab:ex_log}
\end{table}

\subsection{\update{R2.4}{Techniques for Conformance Checking}}
\emph{Conformance checking} is a sub-discipline of process mining, a family of techniques that uses event logs to generate insights for process improvements and to conduct evidence-driven analyses of business processes~\cite{aalst_2012_manifesto}. 
In particular, conformance checking compares traces to the prescribed (to-be) process model to identify deviations between them~\cite{aalst_2022_overview}. 
\update{R2.4}{Consequently, all conformance checking techniques require the same two types of input data: (i) an event log, which captures real-world process executions in the form of traces, and (ii) a process model, which captures the intended process. 
The process model can be understood as a set of rules that the process has to follow, imposed, e.g., by an internal or external regulation or guideline. 
As output, conformance checking techniques identify where and how a trace does not conform with these rules. Based on these individual \emph{violations}, we can compute the \emph{trace fitness}, i.e., the trace's degree of conformance with the rules, and the \emph{log fitness}, i.e., the overall degree of conformance of all traces in the log.

The three most prominent groups of conformance checking techniques are rule-based, token replay-based, and alignment-based \cite{carmona2018conformance}. In the following, we introduce these techniques in detail and explain how they compute violations, trace fitness, and log fitness.}

\update{R2.4}{
\subsubsection{Rule-Based Conformance Checking}
This first group of techniques derives explicit rules from the process model provided as input and checks whether they are fulfilled by individual traces from the event log \cite{carmona2018conformance}. The derived rules can be of different types, including but not limited to:
\begin{compactitem}
    \item Cardinality rules, which define the minimum and maximum number of executions of an activity. For example, the process model in \autoref{fig:ex_model} prescribes that all activities \texttt{A, B, C, D, E,} and \texttt{F} should be executed once at most. 
    \item Response rules, which relate to activity pairs and define that one activity should always follow the execution of the other activity. For example, in \autoref{fig:ex_model}, activity \texttt{D} always has to follow activity \texttt{A}.
    \item Exclusiveness rules, which also relate to pair of activities and define that only one of the activities should be executed, but not both. As an example, consider activities \texttt{E} and \texttt{F} in \autoref{fig:ex_model}.
\end{compactitem}

\noindent
After deriving all rules from a process model, traces are individually checked with respect to these rules. If a rule is not fulfilled, a violation of this specific rule by the given set of activities is identified. Then, trace fitness can be determined by relating the number of violations in a trace to the total number of rules. Log fitness is calculated by taking the average of the fitness values for each trace variant, weighted by their respective frequencies.

\subsubsection{Token Replay-Based Conformance Checking}
This second group of conformance checking techniques tries to execute (replay) each trace in the event log according to the process model~\cite{carmona2018conformance}. 
Note that this is only applicable to models with formal token-based execution semantics, such as BPMN \cite{BPMN}. 
The replay starts by initializing the trace with one token in the initial state of the model, e.g., in the start event in \autoref{fig:ex_model}.
Then, tokens are consumed and produced for each correctly executed activity. 
Whenever a trace does not include an activity required by the model to continue the replay, a missing token is added to the corresponding activity. 
For example, when replaying the trace \texttt{id-7} with activities \texttt{(A,C,B,E)} from \autoref{tab:ex_log} on the process model in \autoref{fig:ex_model}, a missing token would be added to activity \texttt{D} as it is required after activity \texttt{B}.
After the replay, all tokens that have not been consumed are marked as remaining tokens.

Missing and remaining tokens from the token replay are then interpreted as local violations. For example in trace \texttt{id-7}, the missing token at activity \texttt{D} indicates that this required activity is skipped. 
Trace fitness is determined by relating the number of remaining and missing tokens to the total number of consumed and produced tokens during replay. Log fitness is calculated accordingly, based on all missing, remaining, produced, and consumed tokens when replaying all traces in the event log.

\subsubsection{Alignment-Based Conformance Checking}
This third group of techniques constitute the current state-of-the-art for conformance checking \cite{carmona2018conformance}. 
The goal of an alignment is to find an execution sequence of the model that is as close to the trace as possible. 
Compared to token-based replay, this does not rely on token-based semantics, but can be applied to all models that specify execution sequences.
To establish an alignment, we iterate through the activities in trace and search corresponding activities in the process model. If we find one, we add this match to the alignment (synchronous move). If the activity in the trace is not executed in the model, we mark it as inserted (log move). If activities from the model must be executed to continue with the trace, we mark them as missing (model move). 
Both log and model moves are penalized by a cost function, which allows the algorithm to find an optimal alignment for each trace. 
For instance, for trace \texttt{id-7} from \autoref{tab:ex_log}, activity \texttt{D} is missing, so the alignment contains a model move on \texttt{D} in addition to four synchronous moves.

The log and model moves in an alignment represent individual violations in the form of inserted and missing activities.
Trace-level fitness is computed by dividing the cost of the optimal alignment by the cost of the worst-case scenario and deducting the result from 1. 
Log fitness is calculated by taking the weighted average of the fitness values for each trace variant, factoring in their respective frequencies.}


\section{Research Method}
\label{sec:method}

This paper aims to develop a task taxonomy for the conformance checking domain. 
In this section, we explain the research method we followed for that purpose.
In a data analysis context, a task can be understood as a ``question concerning data that can be answered on the basis of the information contained in the data'' \cite[p.646]{andrienko_2006_exploratory}. 
In visualization research, tasks are used to guide the development and evaluation of a visualization system \cite{kerracher_2015_task}. 
By considering \textit{why} users interact with their system, designers can decide \textit{what} data they need and \textit{how} that data should be visualized \cite{munzner2014visualization}.
A task taxonomy supports researchers in obtaining an overview of the multitude of tasks in an analysis domain \cite{munzner2009nested}. 
Such a taxonomy provides a multi-dimensional classification of individual tasks that allow to better relate and structure them \cite{kerracher2017constructing}.
In this paper, we extend the collection of task taxonomies that already exist in the visualization domain (e.g., \cite{kerracher_2015_task,beck2017taxonomy,mcgee2021task}) by developing a task taxonomy for conformance checking.

\update{R2.4}{
It is important to note that we set out to identify tasks in such a way that they were independent from a concrete conformance checking technique. 
As shown in \autoref{sec:background}, existing conformance checking techniques follow the same broad schema: Provided with an event log and a process model, they identify individual rule violations as well as conformance measures for traces and logs. 
This means that from the results alone, it is often not possible to deduce the concrete technique.
However, it also means that in order to identify and classify a task according to our definition above, it is not necessary to consider the technique, as the results suffice to answer the question about the data.
Hence, this approach increases the scope and applicability of our taxonomy to situations where the concrete conformance checking techniques are unknown.
In particular, this is important for commercial process mining tools, who often protect their algorithms as trade secrets. 
}

\begin{figure}[htb]
    \centering
    \includegraphics[width=.7\textwidth]{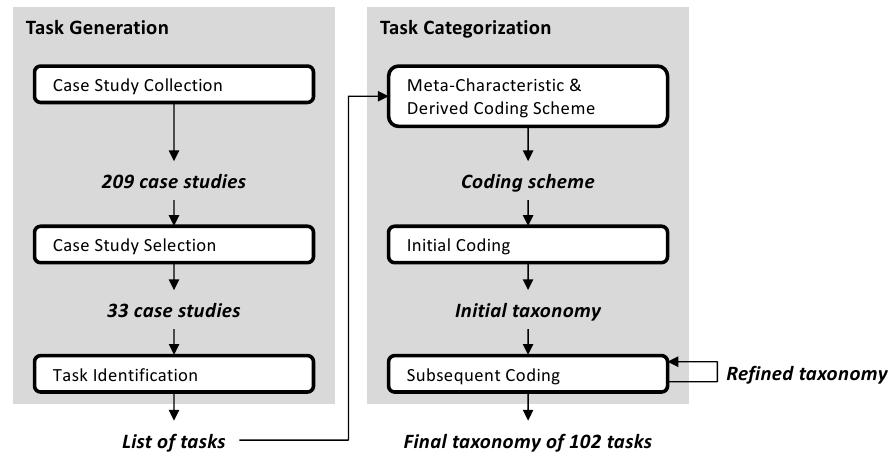}
    \caption{Overview of Applied Research Method}
    \label{fig:method}
\end{figure}

In the following, we describe our method for developing the taxonomy, illustrated in \autoref{fig:method}.
Broadly speaking, it consisted of two steps \cite{kerracher2017constructing}: task generation (\autoref{subsec:generation}), where  we obtained a list of tasks, and task categorization (\autoref{subsec:categorization}), where we structured these tasks into categories. 

\subsection{Task Generation}
\label{subsec:generation}

There are multiple methods to generate tasks, including interviews with domain experts, surveys of visualization experts, observations, or system reviews \cite{kerracher_2015_task}. 
The problem with these methods, however, is that they assume a widespread adoption of the domain and its tasks among practitioners. 
As we have established above, conformance checking is not widely applied in practice~\cite[p.~39]{reinkemeyer2020process}. 
On the contrary, the lack of existing solutions for conformance checking visualization is the main motivation for this paper. 
Therefore, we decided to generate our tasks based on academic literature, where conformance checking is more widely used. 
Specifically, we focused on case studies, i.e., papers where researchers report how they used process mining to answer questions or address problems raised by an (industry) partner \cite{emamjome_2019_case_study}. 
This approach had two advantages: First, due to their inherent technical knowledge, academics are less dependent on the availability of visualizations, meaning that we can expect them to apply conformance checking techniques to their full analytical potential. 
Second, by considering case studies instead of other academic contributions, we can focus on the general capabilities of conformance checking instead of the specifics of individual approaches.

\begin{figure}[htb]
    \centering
        \includegraphics[width=\textwidth]{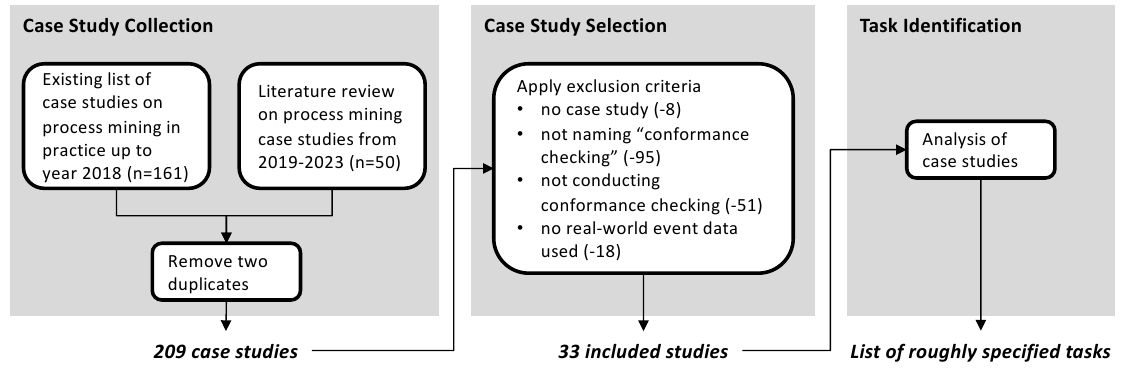}
    \caption{Overview of Conducted Steps in Task Generation}
    \label{fig:generation}
\end{figure}

We conducted a literature review to collect and analyze existing academic case studies, aiming to generate a list of tasks related to the execution and visualization of conformance checking. 
Following the typology of literature reviews~\cite{pare2015synthesizing}, this review combines the descriptive and the critical approach: We first collected conformance checking case studies  and then subjected them to predefined exclusion criteria to ensure that they were relevant to our analysis subject. Finally, we scanned the papers to identify a preliminary set of tasks, which can then be categorized in the next step. \autoref{fig:generation} illustrates these three steps, which are explained in the following sections.

\subsubsection{Case Study Collection}

To identify potentially relevant case studies, we initially relied on an existing review of case studies on process mining in practice \cite{emamjome_2019_case_study}, whose authors kindly provided us with their complete result list of 161 papers.\footnote{Although the original paper \cite{emamjome_2019_case_study} only features 152 process mining case studies, we relied on the provided list, which contained 161 publications.} 
Because they only covered papers up to 2018, we conducted a secondary literature review to also include case studies published between 2019 and 2023 (up to September).
Therefore, we used the search term ``case study'' AND ``conformance checking'' on Google Scholar, limiting the time frame to 2019--2023 and sorting the papers by relevance.\footnote{The queries were run on September 11, 2023.}
Scanning the titles and abstracts of the initial results, we noticed a gradual decline in topical relevance, so we decided to limit ourselves to the 50 most relevant papers from our search. 
Those papers were added to our initial list of 161 papers. 
After removing duplicates (2), our search resulted in a collection of 209 publications.

\subsubsection{Case Study Selection}

To ensure the relevance of the identified literature, we established four inclusion criteria, which defined the characteristics that papers needed to fulfill to be relevant to our research context: 

\noindent
\begin{compactenum}[(1)]
    \item The paper must contain a case study according to our definition above. 
    \item The paper must contain the term ``conformance checking''. 
    \item Within the case study, at least one conducted analysis must make use of conformance checking techniques. It is not sufficient if conformance checking is only applied for the purpose of performance analysis.
    \item Within the case study, conformance checking must be applied on real-world event data. 
\end{compactenum}

\noindent
\update{R2.4}{
We purposefully did not require our case studies to name the concrete conformance checking technique that they applied. 
This aligned with our objective to keep the tasks independent from the techniques. 
In fact, many case studies did not provide detailed insights on how they obtained the presented results, so imposing this criterion would have significantly reduced our results. 
}

We then systematically applied these criteria to select relevant case studies from our collection of 209 publications, as shown in \autoref{tab:criteria}. 
Once a publication did not meet all inclusion criteria, it was excluded from our analysis.
This resulted in 33 case studies, which we found to be suitable to generate a comprehensive list of conformance checking (visualization) tasks.
The final set of case studies can be found online.\footnote{\url{https://figshare.com/s/2c2c505ac602c17d89fe}}

\begin{table}[htb]
\begin{tabularx}{\linewidth}{l|X|c|c}
\multicolumn{2}{l|}{\textbf{Criteria}} & $\Sigma$ & Excl. \\
\hline
\hline
\multicolumn{2}{l|}{Duplicates} & 211 & -2 \\
\hline
(1) & Publications that do not contain at least one case study & 209 & -8 \\
(2) & Publications that do not contain the term ``conformance checking'' & 201 & -95  \\
(3) & Publications that do not conduct conformance checking or limit it to performance analysis & 106 & -55  \\
(4) & Publications that do not apply conformance checking on real-world event data & 51 & -18 \\
\hline
\hline
\multicolumn{2}{l|}{\textbf{Relevant publications}} & \multicolumn{2}{c}{\textbf{33}} \\
\end{tabularx}
\caption{Application of Inclusion Criteria for Case Study Selection}
\label{tab:criteria}
\end{table}

\subsubsection{Task Identification}

In the next step, we analyzed the contents of the 33 case studies in order to identify the conducted tasks. 
For that purpose, we relied on the above-mentioned definition of a task as a question concerning data \cite{andrienko_2006_exploratory}. 
Applied to our domain, we considered a task as an application of an analysis technique, which either answered a concrete question a process stakeholder had previously posed or provided an insight that they did not previously have. 
Here, it was important to limit tasks to the application of a single analysis technique in order to keep them on the same level of abstraction, identify multiple occurrences of the same task, and find dependencies between tasks. 
If the answer to a stakeholder question required the application of multiple analysis techniques, we hence separated it into multiple tasks.

When analyzing the text of a case study paper, there were two ways that we were able to find tasks: 
\noindent
\begin{compactenum}[(1)]
    \item The paper contained a visualization of its analysis result, e.g., in the form of a figure or a table. In this case, we identified one task per visualization, consulting the text to understand the details of the conducted analysis. 
    \item \update{R2.4}{If the paper did not visualize the results of its conducted analysis, we had to rely on the textual description of the conducted analysis. In this case, we identified a task whenever the paper contained a description of a data analysis that was detailed enough to identify the input data and analysis result. Again, we did not require any details on how the result was obtained. For example, it sufficed when a paper stated that they computed a fitness value, the concrete algorithm was not needed.} 
\end{compactenum}

\noindent
In both cases, we identified new tasks when an analysis was set out to answer a previously unanswered question. If two different analyses answered the same questions (e.g., when a case study established the conformance of an event log by means of two different fitness measures), we did not count this as separate tasks, because according to our definition, a task should be agnostic to the underlying analysis technique used to answer the question. 

A few papers also described analyses that the authors would have liked to conduct in order to gain deeper insights into the process but were not conducted due to a lack of available data or suitable analysis techniques. As long as those descriptions were sufficiently detailed, we also included those tasks in our list. The rationale behind this decision was that including those tasks made our list more comprehensive, hence strengthening the validity of our developed taxonomy.

\subsection{Task Categorization}
\label{subsec:categorization}

The task generation step resulted in a list of roughly specified tasks that occurred in at least one of the 33 case studies. 
In the next step, our objective was to use these tasks to develop our actual task taxonomy. 
\update{R1.3}{This taxonomy should classify conformance checking visualization tasks using a set of dimensions, where each dimension consists of at least two mutually exclusive and collectively exhaustive characteristics such that each identified task has exactly one characteristic for each dimension \cite{nickersonMethodTaxonomyDevelopment2013}.} 

To develop the taxonomy, we adopted the iterative development method introduced by Nickerson et al.~\cite{nickersonMethodTaxonomyDevelopment2013}. 
First, we introduced a comprehensive meta-characteristic that guided the choice of characteristics in the taxonomy, and defined ending conditions for the development process. 
After that, the taxonomy was developed iteratively, where each iteration either used an empirical-to-conceptual approach (identifying new tasks and then creating or revising dimensions and characteristics based on them) or a conceptual-to-empirical approach (conceptualizing dimensions and characteristics and then examining tasks for them). 
The taxonomy development was finished if the ending conditions were met after any iteration. 
\update{R2.6}{In our development process, we heavily relied on coding \cite{recker_2021_scientific}, a manual data analysis technique, where raw data is organized into conceptual categories, called codes. Its goal is to identify themes or ideas in heterogeneous data. Each code serves as a higher-level category under which a piece of data can be subsumed. Specifically, we employed an open coding technique, with the goal of covering our full dataset in a set of codes.}

\begin{figure}[htb]
    \centering
    \includegraphics[width=.9\textwidth]{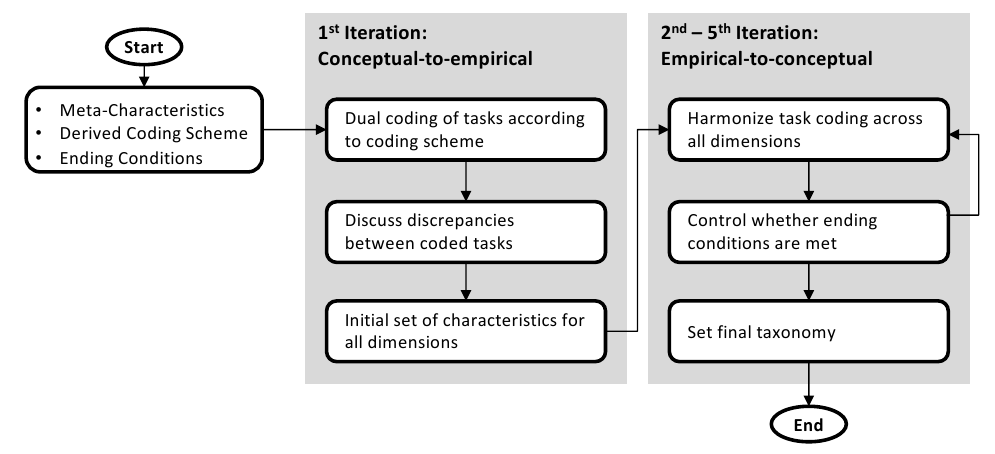}
    \caption{Overview of Conducted Steps in Task Categorization}
    \label{fig:taxonomy_iterations}
\end{figure}

\update{R2.5}{
The iterative development process for our taxonomy is illustrated in \autoref{fig:taxonomy_iterations}.
For our meta-characteristic, we relied on the design space of visualization tasks  to derive a generic coding scheme, as described in \autoref{subsubsec:meta}. 
Our first iteration was conceptual-to-empirical, where we categorized the identified tasks according to this scheme, as described in \autoref{subsubsec:initial}. 
The remaining iterations were empirical-to-conceptual, where we inductively refined our task list and coding scheme to be more suitable to our application domain, as described in \autoref{subsubsec:subsequent}.
Our ending conditions, adopted from Nickerson et al.~\cite{nickersonMethodTaxonomyDevelopment2013}, were met after the fifth iteration. 
}

\subsubsection{Meta-Characteristic \& Derived Coding Scheme}
\label{subsubsec:meta}

\update{R2.7}{The meta-characteristic defines the purpose of the taxonomy and serves as a initial orientation for selecting its dimensions and characteristics. 
For our taxonomy, the meta-characteristic should hence provide a generic understanding of tasks, which we could then adapt to conformance checking. 
Therefore, we selected the design space of visualization tasks~\cite{schulz2013design,tominski2020interactive} as a meta-characterics from which to derive our initial dimensions and characteristics. 
This design space is widely used to characterize visualization tasks. 
To provide a comprehensive perspective on a task, it encompasses five dimensions: \textit{task goal}, \textit{task means}, \textit{data characteristics}, \textit{data target}, and \textit{data cardinality}. In the following, the dimensions are explained in more detail~\cite{schulz2013design,tominski2020interactive}.}

\paragraph{Task Goal} 
This dimension specifies the purpose behind a task (Why does a user pursue this task?), delineating the motive behind its actions~\cite{schulz2013design,tominski2020interactive}. In our initial coding scheme, this dimension contained five characteristics, i.e.,  potential task goals: \textit{Explore}, \textit{Describe}, \textit{Explain}, \textit{Confirm}, or \textit{Present}.

\paragraph{Task Means} 
This dimension specifies the methods employed to achieve the task goal (How is the task carried out?). It is independent of the task goal, as the same goal can be achieved through different means and vice versa~\cite{schulz2013design,tominski2020interactive}. 
To define the initial characteristics for this dimension, we referred to the comprehensive framework by Munzner~\cite{munzner2014visualization}, which provides a frequently-cited topological classification of task means (or actions). 
The framework separates the potential tasks means into three high-level groups~\cite{munzner2014visualization}: 
Task means in the ``analyze'' group include, for example, \textit{discover} (finding new, previously unknown knowledge) and \textit{annotate} (manually adding graphics or texts to an existing visualization).
The ``search'' group includes, for example, \textit{locate} (finding a known target in an unknown location) and \textit{browse} (finding an unknown target in a known location).
Finally, the ``query'' group includes, for example, \textit{identify} (specifying the exact characteristics of a single target) and \textit{summarize} (providing an overview over all targets). 
For our purposes, we used the lower-level task means as they were the most specific. 

\paragraph{Data Characteristics} 
\update{R2.7}{This dimension specifies the facets of the data that the task should reveal (What does a task seek?)  \cite{schulz2013design,tominski2020interactive}. 
Due to the abundance of (visualized) data, there are many different aspects that influence the characteristics of a task and its data \cite{gleicher2023problem}, which is why generic visualization frameworks like the design space can only provide a very general classification for this dimension, for example by differentiating between low-level and high-level data characteristics \cite{schulz2013design,tominski2020interactive}. 
For our taxonomy, however, we wanted the data characteristics to be more specific and focused on the conformance checking domain, i.e., related to the potential outcomes of a conformance checking analysis. 
For this reason, we did not pre-define which data characteristics we expected in our conformance checking tasks, like we did for the previous dimensions. 
Instead, we went into the initial coding without pre-defined codes and inductively identified the characteristics during the coding process. This means that, after starting with an empty set of codes, we created a new one when we came across a new data characteristic that did not fit our existing set of codes.}

\paragraph{Data Target}
\update{R2.7}{This dimension, together with the next one, specifies the data on which the task is conducted (Where in the data is the task carried out?) \cite{schulz2013design,tominski2020interactive}. 
Concretely, the data target identifies the part of the data on which the task is carried out.  
This dimension was similar to the previous one, in the sense that we wanted the data targets to be specific to the conformance checking domain, which typically involves data in the form of events logs and process models. 
Hence, we employed the same strategy as for the data characteristics: We went into the initial coding without a set of pre-defined codes for the data targets and employed an inductive strategy to directly derive the codes from our data.} 

\paragraph{Data Cardinality}
This dimension specifies how many instances of the chosen data target are considered by a task, complementing the data target \cite{schulz2013design}. For our initial coding scheme, we adopted the characteristics proposed by the design space of visualization tasks~\cite{schulz2013design,tominski2020interactive}, who differentiate between \textit{single}, \textit{multiple}, and \textit{all} instances.

\bigskip 
\noindent
This initial set of dimensions and characteristics served as the basis for our iterative coding, as described next. 
Before starting the iterations, we chose the objective ending conditions from Nickerson et al.~\cite{nickersonMethodTaxonomyDevelopment2013}: All tasks have been examined, no tasks, dimensions, or characteristics were changed in the last iteration, and every dimension and characteristic is unique. 

\subsubsection{Initial Coding}
\label{subsubsec:initial}

In our first, conceptual-to-empirical, coding iteration, the goal was to specify conformance checking tasks as individual five-tuples, according to the five presented dimensions of the design space \cite{schulz2013design}.
To ensure coding reliability, we employed a double coding approach. Each of the 33 case studies was assigned to two randomly selected authors of this paper, who independently categorized the rough conformance checking tasks identified in the previous section. Discrepancies in task categorization were discussed among all authors until an unambiguous five-tuple was defined for each task. 

In a first step, this coding strategy was applied to ten randomly selected case studies.
This allowed us to develop an initial standardized vocabulary for our domain, thereby particularly focusing on the design space dimension of data characteristics and data target, where we did not have any predefined characteristics.
For example, we realized that in order to stay technology-agnostic, it was better to denote the specifications of prescribed process behavior as  ``guidelines'' instead of the more common, but not technology-agnostic ``process model''.  
This became evident after we found some case studies that conducted conformance checking by means of rule checking approaches \cite{carmona2018conformance,BURATTIN2016194}, which rely on a set of rules instead of a full process model to specify the prescribed process behavior. 

Using the established vocabulary, the double coding procedure was applied to all remaining case studies. Again, the categorizations by the two assigned coders were compared and discussed among all authors until full agreement was achieved. 
The result of this initial coding was a list of specified tasks, which could be further refined in subsequent coding iterations.

This iteration also established an initial set of characteristics for our two inductively defined dimensions, data characteristics and data target. 
With regard to the latter, we found that due to the predefined input of conformance checking (an event log and a set of guidelines), the data target mainly varied in the granularity of its analysis. 
We identified three levels of granularity. Tasks either referred to (a) the full \emph{log}, e.g., when calculating the conformance measure of a process, (b) an individual \emph{trace}, e.g., when identifying concrete violations of the guidelines, or (c) one or multiple \emph{events}, e.g., when investigating patterns of violations. 

The data characteristics dimension was more complicated. 
We found a small set of frequently occurring characteristics, which typically referred to established concepts in conformance checking, such as the overall \emph{conformance of a log} (assessed in terms of a fitness measure).
In addition, we found a large number of characteristics that only occurred a few times, such as the \emph{severity of guideline violations}. 
This long-tail distribution of data characteristics showed the variability of the conformance checking domain, where individual analysis questions sometimes require specific techniques. It also indicates that many analyses do not tap the full potential of conformance checking, which underlines the value of our proposed taxonomy.

\subsubsection{Subsequent Coding Iterations}
\label{subsubsec:subsequent}

After the initial coding, we iterated through our initial lists of tasks to harmonize and refine them. 
In total, we went through four additional coding iterations until our ending conditions were met.

\paragraph{Iteration 2: Harmonize within Task Goal}
In this iteration, we separated the tasks per task goal and ensured that within each task goal, the tasks were categorized consistently. 
This was particularly important for the inductively defined data characteristics. 
The iteration mainly contained two measures:
\begin{compactitem}
    \item We removed synonyms from the data characteristics, such that the same entities were denoted by the same words. For example, we considered ``guideline violation'' to be synonymous to ``guideline deviation''.
    \item We also removed homonyms from the data characteristics, such that different entities were denoted by different words. For example, the severity of a guideline violation is not the same as the impact of a violation on the process outcome. 
\end{compactitem}

\paragraph{Iteration 3: Harmonize across Task Goals}
In the next iteration, we harmonized the task categorizations across the task goals, which was easier to accomplish due to the initial harmonization in the previous iteration. 
This iteration mainly contained two measures:
\begin{compactitem}
    \item We harmonized the assigned task goal and task means characteristics to match their original meaning from visualization literature. 
    For example, we had first categorized the computation of a process conformance measure under the task means ``summarize'', but then learned that ``derive'' was better suited because this computation resulted in new knowledge about the process and its degree of conformance. 
    \item We ensured that all our characteristics, particularly the data characteristics, used technology-agnostic language. For example, we named the task ``derive process conformance'' instead of the more technical ``derive fitness measure''. We also removed all references to specific conformance checking techniques, such as alignments.
\end{compactitem}
Once we had harmonized the data characteristics, we realized that many tasks only differed in terms of the process perspective that they referred to. For example, it was possible to list conformant and non-conformant traces both in terms of the data perspective (violating guidelines on the involved data) and the resource perspective (violating guidelines on the involved resources). 
To account for these similarities, we introduced the \textit{constraint type} as a sixth dimension in our taxonomy, defined by the four characteristics \textit{control-flow}, \textit{data}, \textit{resource}, and \textit{time}. As we can see, these correspond to the perspectives typically captured in process models. 
The constraint type dimension allowed us to simplify the data characteristics and hence to immediately identify tasks that occurred in multiple case studies.
To ensure mutually exclusivity in the constraint type, the characteristics are defined as subsets of a set containing all four process perspectives. For example, a task can refer to the \textit{control-flow} and \textit{data} perspective and thus is characterized by this subset of process perspectives in the constraint type.

\enlargethispage{\baselineskip}
\paragraph{Iteration 4: Describe Tasks}
To ensure that our tasks were sufficiently distinct from one another and that the characteristics were unique, we wrote short textual descriptions of each task. 
In this process, we noticed that some tasks differed only in small details, so it made sense to merge them into one slightly more abstract task.
For example, we abstracted the data target from ``Describe: Present distribution over time'', such that it could be merged with ``Describe: Present distribution''. 
For other tasks, we noticed that although they sound similar, they refer to different analyses. For example, ``Describe: Derive process conformance'' denotes the overall degree of process conformance expressed by a fitness value, whereas ``Describe: Summarize process conformance'' denotes the percentage of traces in which a violation occurs.

\paragraph{Iteration 5: Control}
In our final iteration, we inspected our task categorizations once more to ensure that our ending conditions were met. This iteration resulted in our final list of 102 tasks, which are described in the next section. 
The final taxonomy and the coding of all 102 tasks can be found online\footnote{\url{https://figshare.com/s/2c2c505ac602c17d89fe}}.

\section{The Taxonomy}
\label{sec:taxonomy}

\begin{figure}[htb]
    \centering
    \includegraphics[width=\textwidth]{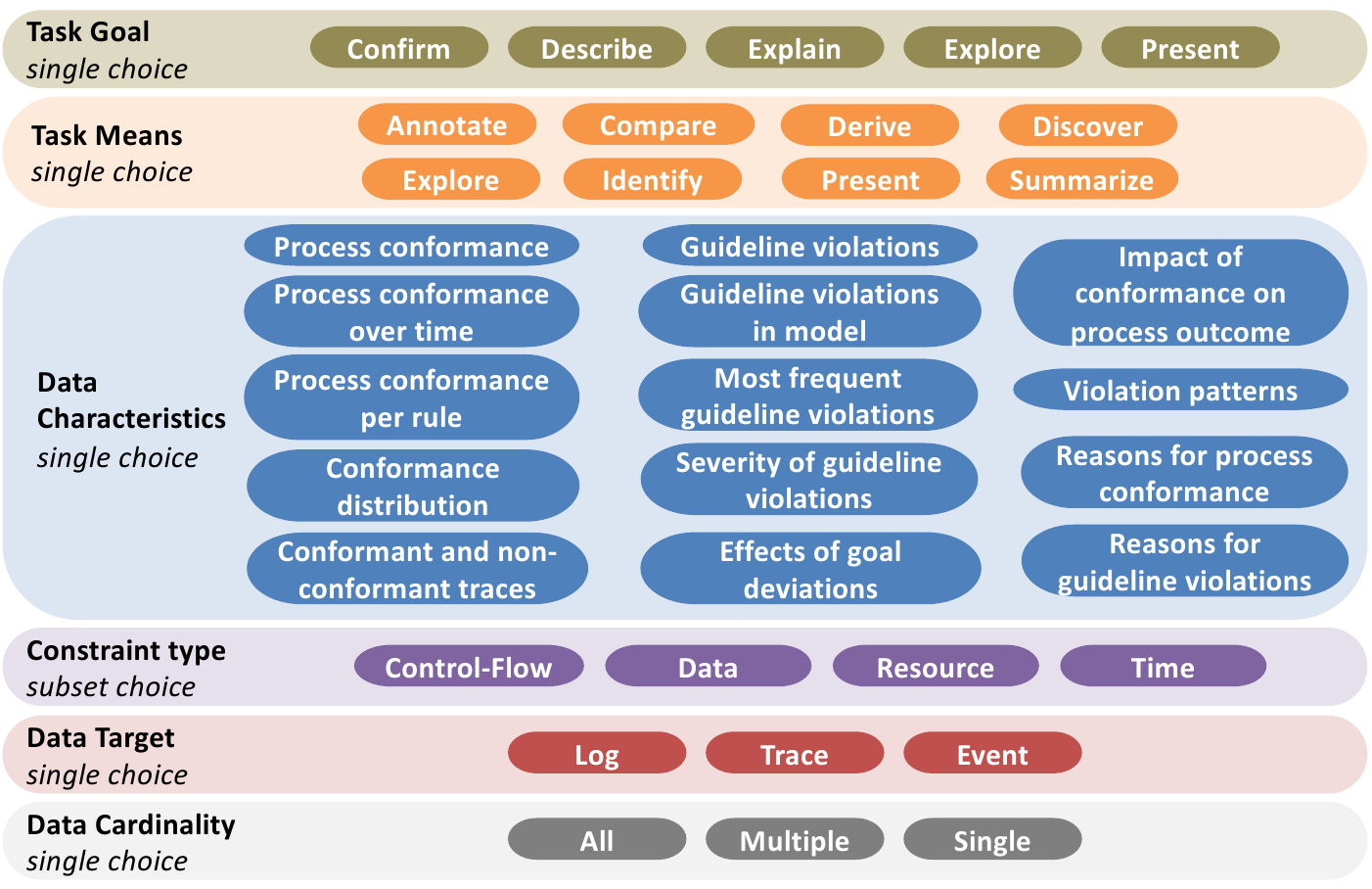}
    \caption{Taxonomy of Conformance Checking Tasks}
    \label{fig:tax}
\end{figure}

The final task taxonomy for conformance checking visualization consists of the six dimensions \textit{task goal}, \textit{task means}, \textit{data characteristics}, \textit{constraint type}, \textit{data target}, and \textit{data cardinality}. These dimensions and all corresponding characteristics that occurred in our visualization tasks are illustrated in \autoref{fig:tax}. 
In the following, we explain the six dimensions, provide individual descriptions of the 102 tasks, and investigate the relations between them. 

\subsection{Task Dimensions}

\begin{figure}[htb]
    \centering
    \includegraphics[width=\textwidth]{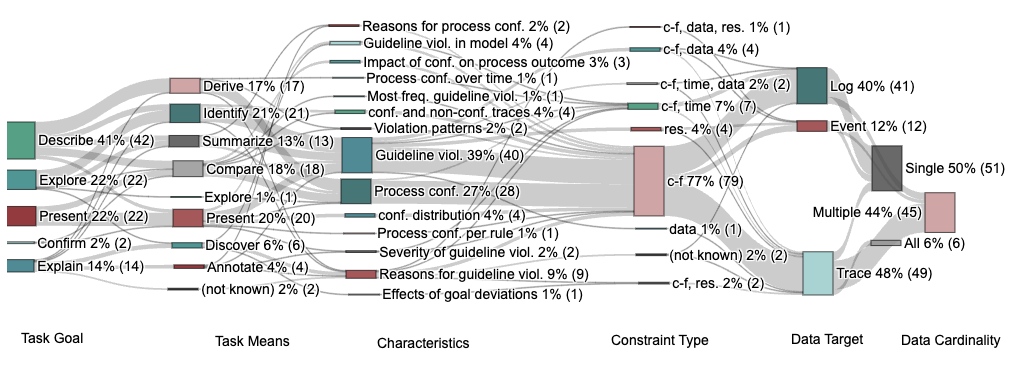}
    \caption{Sankey Diagram of all Realizations in the Task Taxonomy (abbreviations: control-flow (c-f), resource (res.), violations (viol.), conformance (conf.))}
    \label{fig:sankey}
\end{figure}

\begin{figure}[ht]
\centering
\begin{subfigure}{.49\textwidth}
    \centering
    \includegraphics[width=\textwidth]{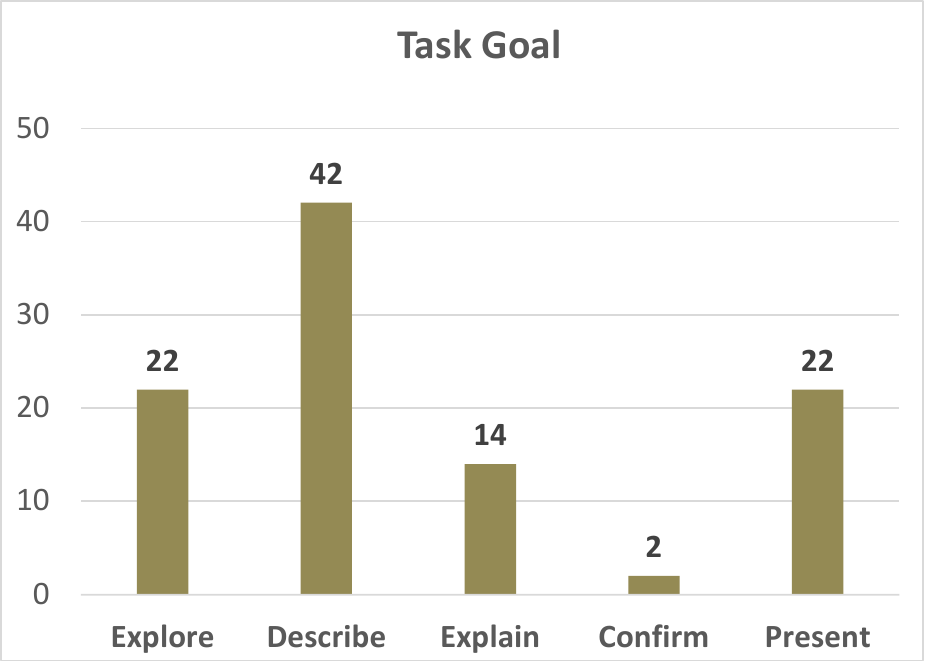}
    \caption{Task Goal}
    \label{fig:goal}
\end{subfigure}
\begin{subfigure}{.49\textwidth}
\centering
    \includegraphics[width=\textwidth]{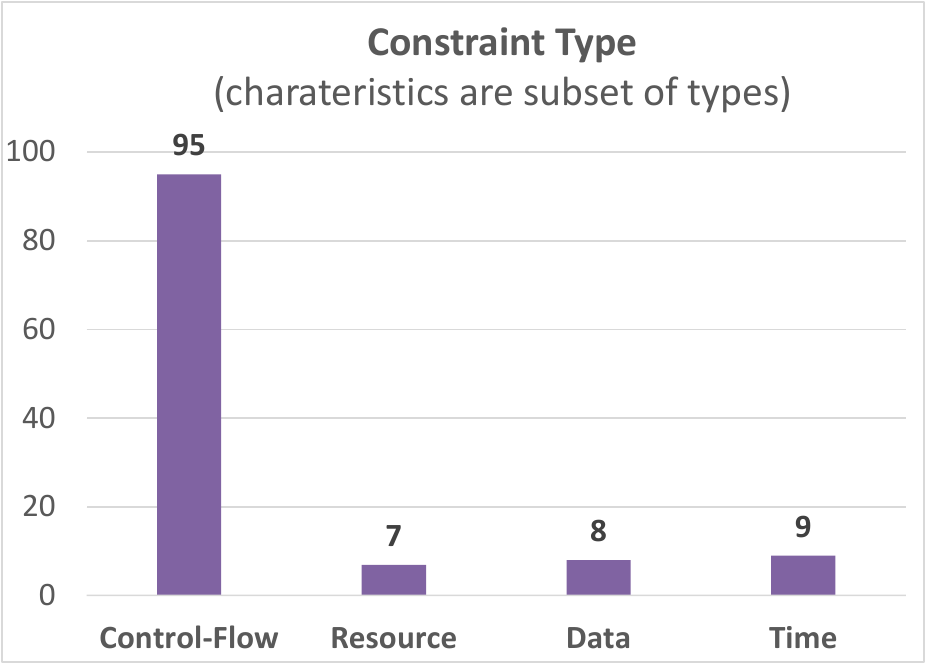}
    \caption{Constraint Type}
    \label{fig:type}
\end{subfigure}
\begin{subfigure}{.49\textwidth}
    \centering
    \includegraphics[width=\textwidth]{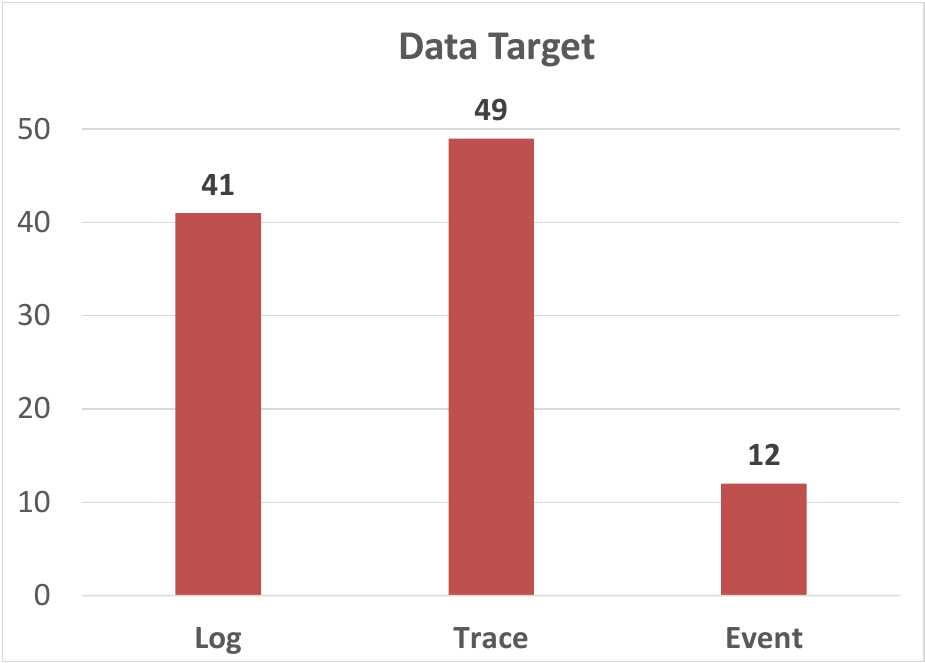}
    \caption{Data Target}
    \label{fig:target}
\end{subfigure}
\begin{subfigure}{.49\textwidth}
\centering
    \includegraphics[width=\textwidth]{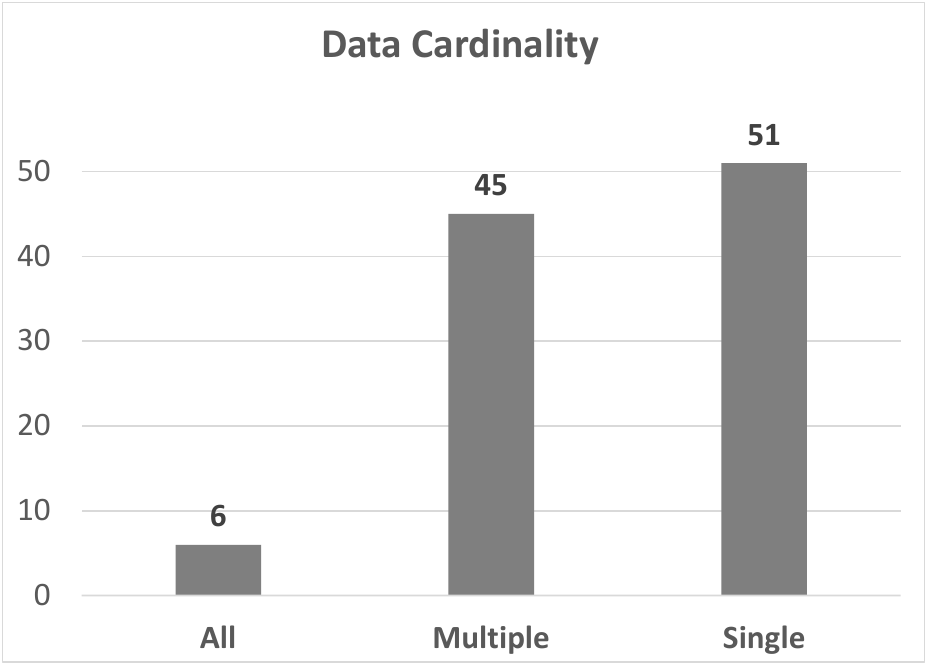}
    \caption{Data Cardinality}
    \label{fig:cardinality}
\end{subfigure}
\caption{Realization of Coding Dimensions of all 102 Tasks} \label{fig:distribution}
\end{figure}

In this section, we provide an overview of how the six dimensions of the taxonomy are distributed over the tasks. \autoref{fig:sankey} shows a Sankey diagram with the realizations of all tasks within the taxonomy characteristics as well as their connections.
Four dimensions have five or fewer characteristics that cover all tasks (task goal, constraint type, data target, and data cardinality). The distribution of characteristics of those four dimensions over all tasks is shown in \autoref{fig:distribution}.\footnote{Note that the constraint type does not add up to the total number of tasks (102) because the characteristics of this dimension are mutually exclusive subsets of the four process perspectives.} 
The remaining two dimensions (task means and data characteristics) have 8 resp. 14 different characteristics, listed in \autoref{fig:top5}. This indicates that they are the most interesting ones when analyzing process conformance, also allowing for variations in the other dimensions. 

\begin{table}[thb]
\centering
\caption{Realizations of Task Means and Characteristics across all 102 Tasks} \label{fig:top5}
\resizebox{\textwidth}{!}{
\begin{tabular}{L{.7cm}lR{1.2cm}|| L{.7cm}L{7.9cm}R{1.2cm}}
\toprule
\multicolumn{3}{l||}{Task   Means} & \multicolumn{3}{l}{Data Characteristics} \\ \midrule
 & Realization & Count &  & Realization & Count \\
1 & Identify & 21 & 1 & Guideline violations & 40 \\
2 & Present & 20 & 2 & Process conformance & 28 \\
3 & Compare & 18 & 3 & Reasons for guideline violations & 9 \\
4 & Derive & 17 & 4 & Guideline violations in model & 4 \\
5 & Summarize & 13 & 5 & Conformance distribution & 4 \\
6 & Discover & 6 & 6 & Conformant and non-conformant traces & 4 \\
7 & Annotate & 4 & 7 & Impact of conformance on process outcome & 3 \\
8 & Explore & 1 & 8 & Reasons for process conformance & 2 \\
 &  &  & 9 & Severity of guideline violations & 2 \\
 &  &  & 10 & Violation patterns & 2 \\
 &  &  & 11 & Effects of goal deviations & 1 \\
 &  &  & 12 & Process conformance per rule & 1 \\
 &  &  & 13 & Process conformance over time & 1 \\
 &  &  & 14 & Most frequent guideline violations & 1 \\ \bottomrule
\end{tabular}}
\end{table}

We see a predominant focus on the task goals \textit{describe}, \textit{explore}, and \textit{present}, which occur in 86 tasks. 
Conversely, \textit{explain} and \textit{conform} rarely occur: Only 14 tasks explain non-conformance and only two confirm a hypothesis about the process conformance. This supports the observation that process mining in general predominantly follows an exploratory instead of a confirmatory paradigm \cite{janssenswillen_2019_confirmatory}.
The constraint type shows a very strong focus on the control-flow of the process. This also supports findings on the focus of conformance checking in general~\cite{dunzer_2019_conformance_checking}. 

For the data targets, log and trace are used most frequently, but only few tasks target conformance at the event level. This indicates that a broader overview is provided more often than a fine granular analysis. As for the data cardinality, rather few tasks aim to visualize all targets, but most focus on multiple or single targets. This makes sense when remembering the hierarchical structure of an event log, where logs consist of traces and traces consist of events. If a task concerns, e.g., all events of a trace, it more likely refers to the trace level, such that it will be coded as ``trace (single)'' instead of ``event (all)''. One notable exception to this is the task ``Describe: Present conformance distribution'', which refers to the conformance of individual traces, but presents their distribution, e.g., over time. 

In \autoref{fig:top5}, we see that the top five realized characteristics of the task means dimension account for 83\% of the tasks.\footnote{Note that the realized characteristics of the task means dimension do not sum up to 102 as the task means is unknown for two tasks.} 
These five realizations are all used at least 13 times, with the identification of conformance issues being the most frequent one. After that, more static means like present and compare dominate over more interactive means like discover, which occurs less frequently. 
The data characteristics show a very skewed long-tail distribution over the tasks. Concretely, the two characteristics \textit{Guideline violations} and \textit{Process conformance} are used in 67\% of the tasks, whereas ten other characteristics are used at most four times.
This indicates the high variability in conformance checking tasks, which holds potential for the development of new visualization systems. 

\subsection{Individual Task Descriptions}
Based on the task taxonomy, the conformance checking visualization tasks can be characterized across the six dimensions. Thereby, each task is a 6-tuple, consisting of one realized characteristic of each dimension. Only for the constraint type, tasks can be associated to multiple realizations.
\autoref{fig:top8tasks} shows the eight most frequently occurring tasks. 
In the following, we describe these eight tasks and show an exemplary visualization found in the case studies, where appropriate. 
The remaining task descriptions can be found online in our repository.

\begin{table}[htb]
\centering
\caption{Top 8 Realizations of Tuple ``Task Goal: Task Means, Data Characteristics''} \label{fig:top8tasks}
\begin{tabular}{lll}
\toprule
\centering
No. & Realization & Count \\ \midrule
1 & Describe: Derive Process conformance & 14 \\
2 & Present: Present Guideline violations & 10 \\
3 & Explore: Identify Guideline violations & 10 \\
4 & Describe: Compare Process conformance & 7 \\
5 & Describe: Identify Guideline violations & 7 \\
6 & Explore: Summarize Guideline violations & 6 \\
7 & Explain: Discover Reasons for guideline violations & 4 \\
8 & Describe: Present Conformance distribution & 4 \\ \bottomrule
\end{tabular}
\end{table}

\paragraph{Describe: Derive Process conformance (control flow, log, single)} 
What is the overall degree of conformance between a (single) log and a set of guidelines? 
The most frequent occurring task quantifies this overall degree of conformance, typically expressed as an automatically computed fitness value between 0 (no conformance at all) and 1 (perfect conformance). This tasks often serves as a first step in a conformance checking analysis, providing a first orientation on which steps to take next. It provides broad feedback on the process and always occurs in conjunction with the constraint type ``control-flow'', the data target ``log'', and the data cardinality ``single''.

\begin{figure}[ht]
    \centering
    \includegraphics[width=\textwidth]{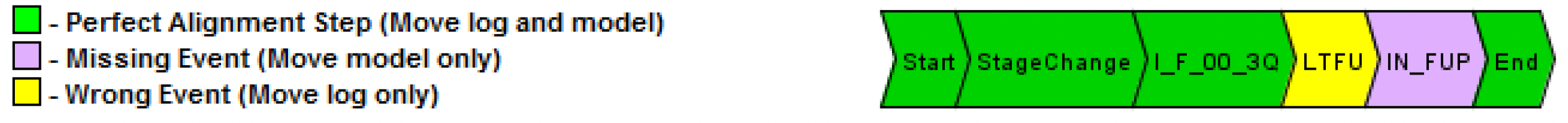}
    \caption{Typical Visualization for ``Present: Present Guideline violations'' \cite{rinner2018process}}
    \label{fig:chevron}
\end{figure}

\paragraph{Present: Present Guideline violations (control flow/data, trace, single/multiple)} 
Where does the recorded behavior violate which guidelines? This is typically answered for one trace or a few traces simultaneously. Concretely, we are interested in where the activities in the trace(s) deviate from the desired process. That way, violations can be attributed to activities in the trace(s). Although the control flow perspective occurs most frequently, violations can also concern the data perspective. The most prominent visualization for this task (referring to control flow, trace, single) uses chevron diagrams, as illustrated in \autoref{fig:chevron}. One chevron diagrams represents one trace under consideration, with colors indicating conforming events (here: green), missing events (here: purple), and wrongly executed events (here: yellow). Typically, this visualization relies on alignments as the underlying technique that classifies events into these three categories.

\paragraph{Explore: Identify Guideline violations (control flow/resource/data, log/trace/ event, single/multiple)}
Where exactly does the process execution differ from the guideline? What does alternative behavior look like? This can relate to different control-flow relations, but also resource and data constraints. Although most frequently relating to traces, it can also occur on an event or log level. The task is similar to ``Present: Present Guideline violations''. The difference is that in this task, the violations are not immediately shown to the analysts. Instead, the means are provided for the analysts to explore the data and identify the violations on their own. 

\paragraph{Describe: Compare Process conformance (control flow, log/trace, single/multiple)}
How does the overall degree of conformance with a set of guidelines differ between multiple logs or traces? To answer this, we first need to derive the process conformance for these entities. At the log level, this provides a more generic overview between different process executions. At the trace level, this is specific to the execution patterns found in the trace. In most cases, these derived values of process conformance are compared as natural numbers.

\paragraph{Describe: Identify Guideline violations (control flow/resource/data, log/trace /event, single/multiple)}
How exactly does the process execution differ from the guidelines? What behavior was prescribed by the guidelines and what behavior was executed instead? This can relate to different control-flow relations, but also resource and data constraints. 
This task is similar to ``Present: Present Guideline violations'', but rather than analyzing the control-flow of a complete trace, ``Describe: Identify Guideline violations'' provides detailed context about specific constraints of interest and their potential violations.

\paragraph{Explore: Summarize Guideline violations(control-flow, log/trace/event, single/multiple)}
What type of guideline violation happened in different traces? How often do they happen? The task is similar to ``Describe: Summarize Guideline violation'', with the difference that ``Explore'' requires the process analyst to summarize the violations on their own.

\begin{figure}[htb]
    \centering
    \includegraphics[width=0.8\textwidth]{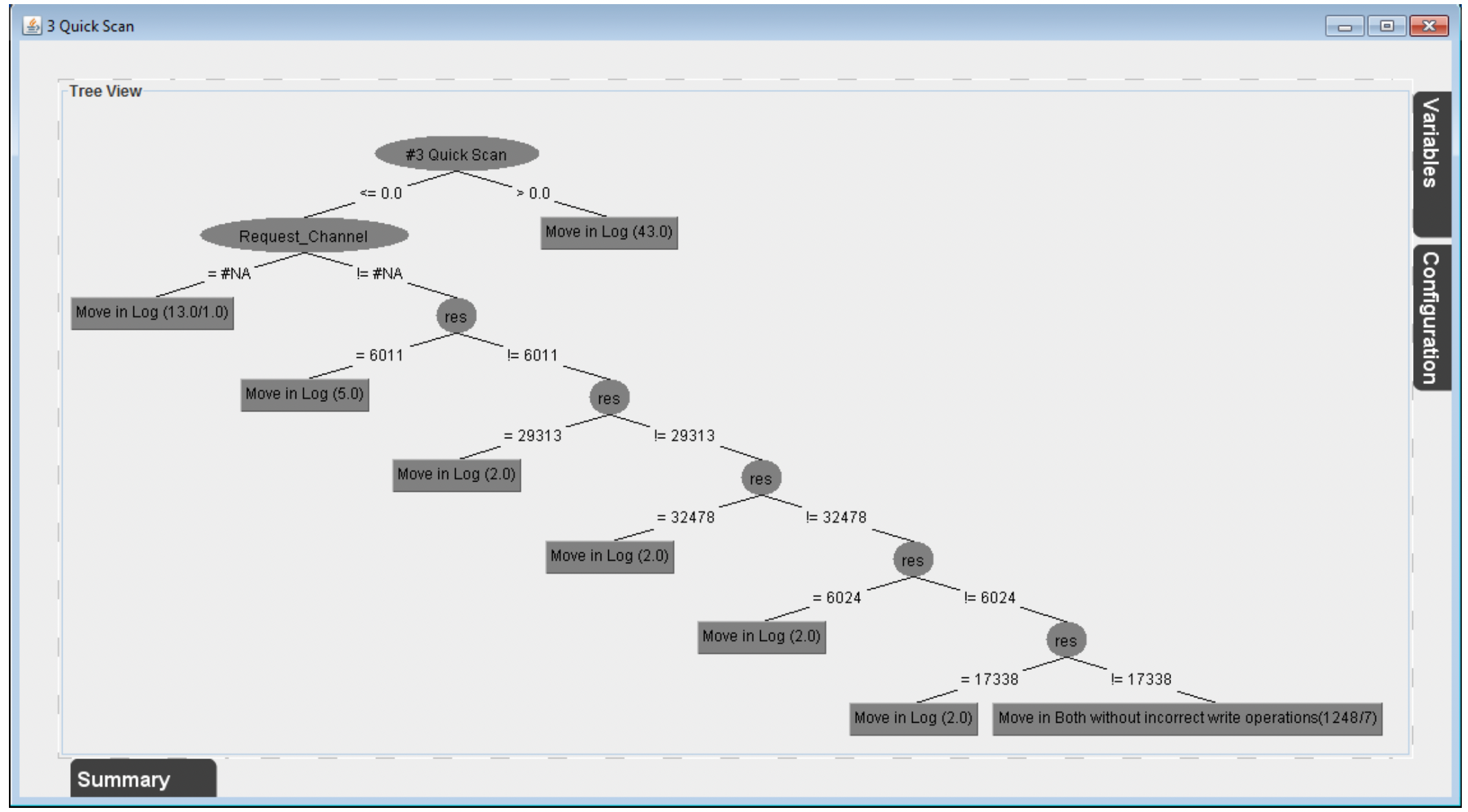}
    \caption{Typical Visualization for ``Explain: Discover Reasons for guideline violations'' \cite{leoni2013aligning}}
    \label{fig:explanation}
\end{figure}

\paragraph{Explain: Discover Reasons for guideline violations (control-flow/resource/ data/time, log/trace/event, single/multiple)}
What control-flow, data, resource, or time attributes of events, traces, or logs lead to guideline violations? Several variations in attributes can be utilized to find statistical evidence for reasons of guideline violations. One prominently used technique for that are regression or decision trees including the respective visualizations, as illustrated in \autoref{fig:explanation}.

\begin{figure}[htb]
    \centering
    \includegraphics[width=0.8\textwidth]{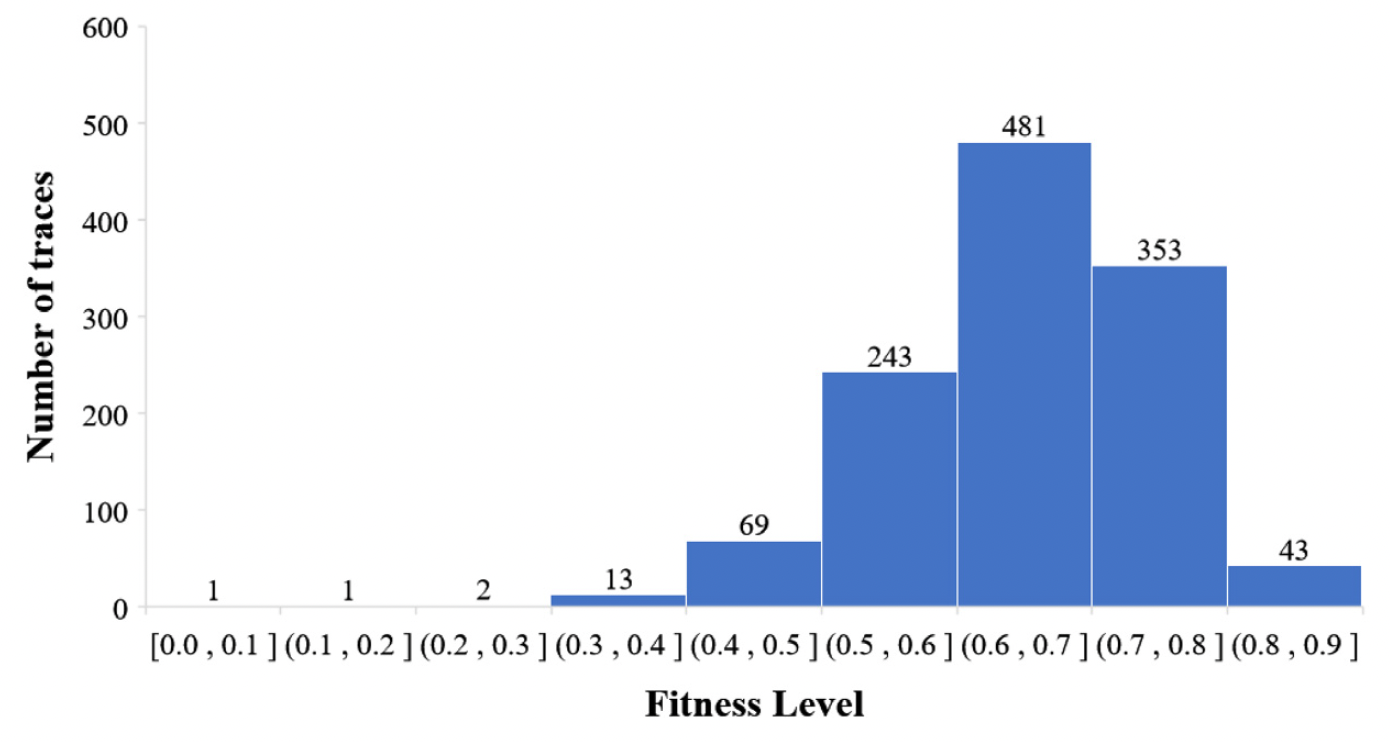}
    \caption{Typical Visualization for ``Describe: Present Conformance distribution'' \cite{zhang2022fuzzy}}
    \label{fig:conf_dist}
\end{figure}

\paragraph{Describe: Present Conformance distribution (control-flow, trace, multiple/all)}
Which percentage of traces in the log fall into which conformance category? In this context, the conformance range is typically separated into intervals, into which the traces are distributed. It allows for a more detailed insight into the overall log conformance, which is an average value. As an example, consider the distribution in \autoref{fig:conf_dist}.

\subsection{Dependencies between Tasks}

After describing the task taxonomy and showing how the realizations of the conformance checking tasks are distributed across the six dimensions, we now analyze dependencies between the tasks. 
Such dependencies potentially unravel inherent workflows in conformance checking analyses and help to better understand the corresponding visualization requirements.
They describe potential paths that a user can take through the system: If one task can only be started once another one has been completed, this dependency has to be considered in the system. 

To identify dependencies between conformance checking tasks, we assess whether certain tasks often follow each other, indicating that it is necessary or just helpful to perform one task before the other. 
For that purpose, we make use of process mining methods, concretely process discovery. The idea is that we consider conformance checking analyses as a process, where our tasks are the activities. Each of our case studies is one execution of that process, i.e., one trace. If we discover patterns in the control flow of that process, this means that the same set of tasks were executed in the same order in different case studies, which hints at a dependency between the tasks. 

\begin{figure}[htb]
    \centering
    \includegraphics[width=\textwidth]{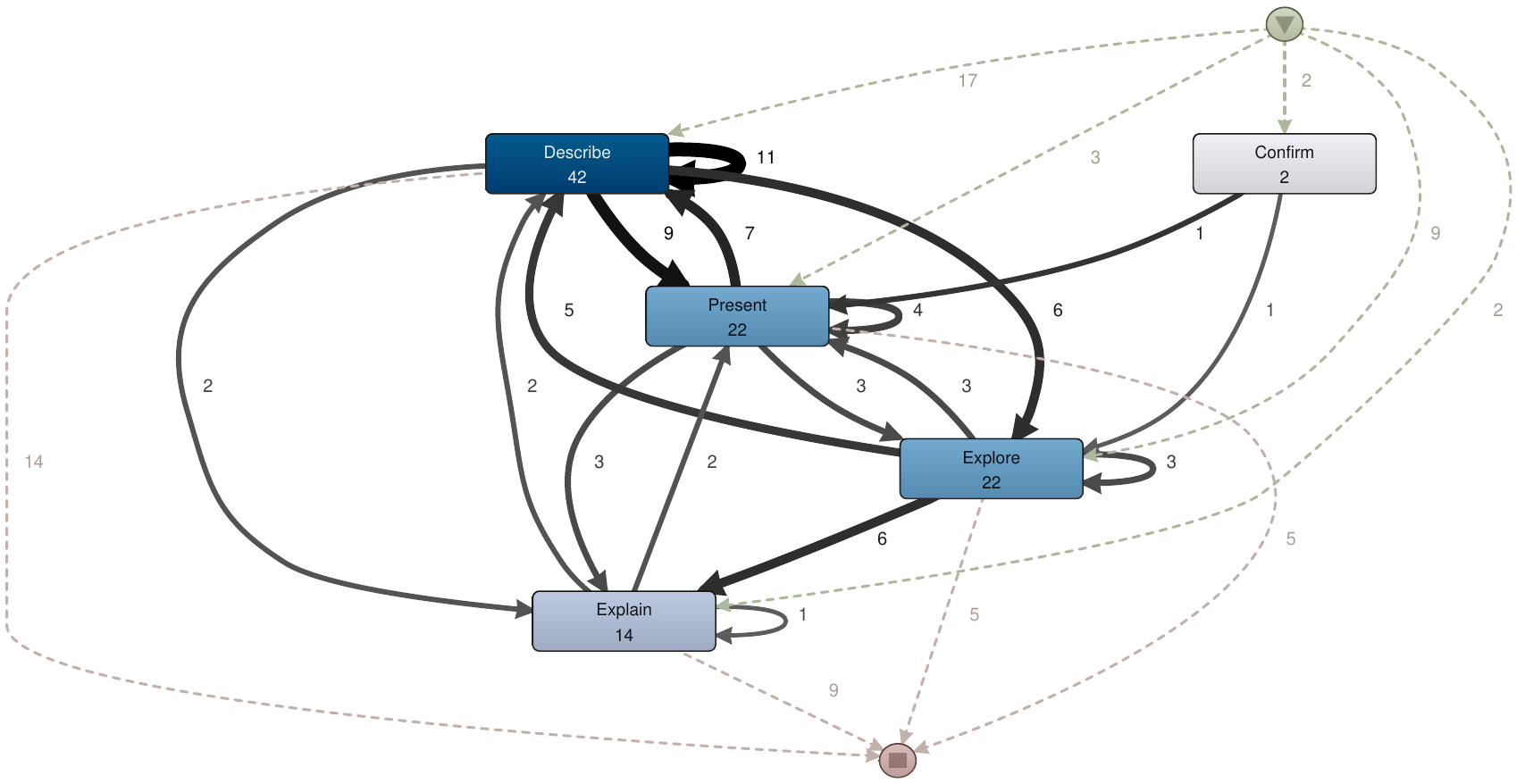}
    \caption{Discovered Model for Tasks -- Activities defined by Task Goal}
    \label{fig:task_goal}
\end{figure}

\begin{figure}[htb]
    \centering
    \includegraphics[width=.8\textwidth]{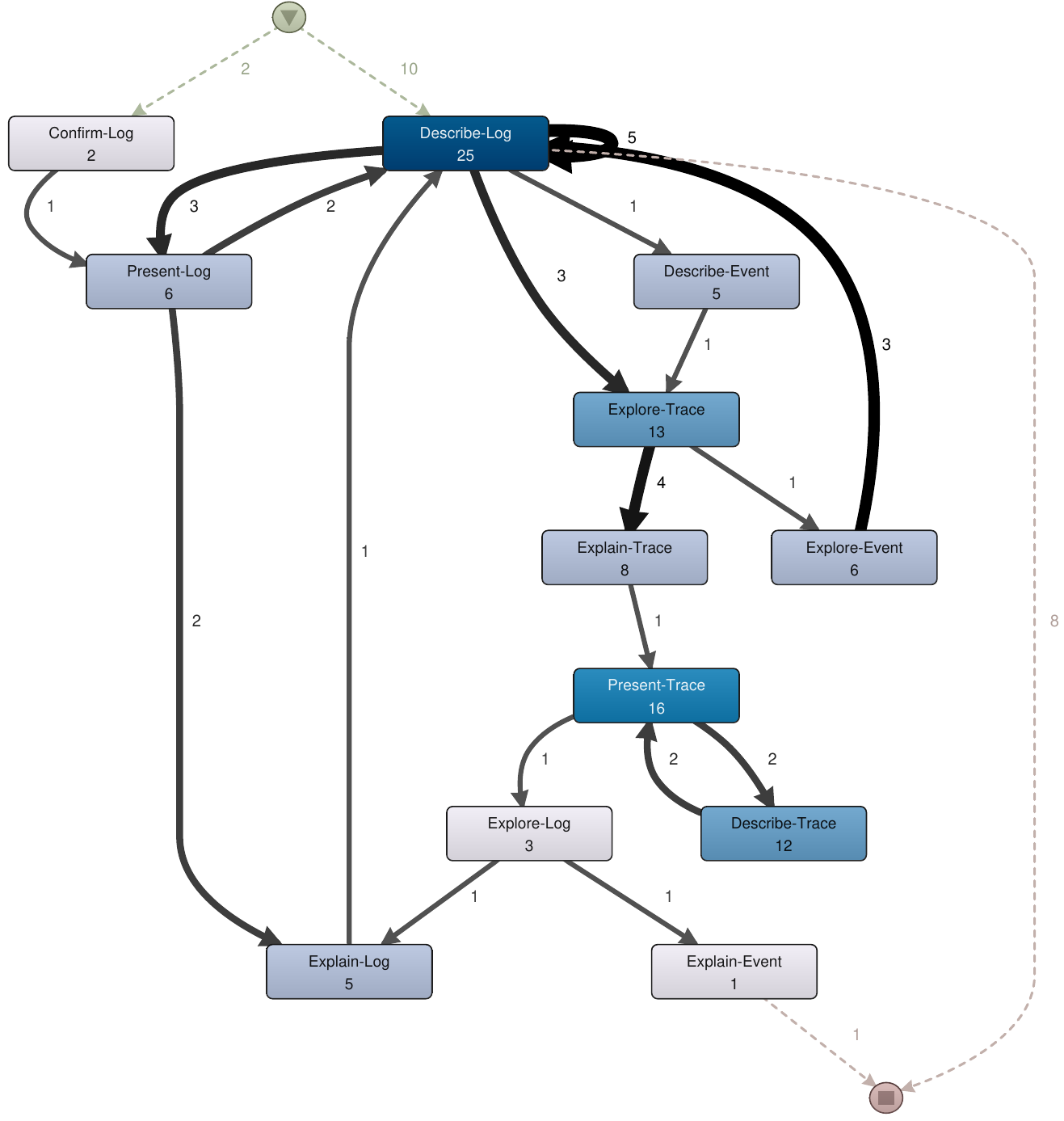}
    \caption{Discovered Model for Tasks -- Activities defined by Task Goal and Data Target (18\% paths)}
    \label{fig:task_goal_target}
\end{figure}

To create an event log from our case studies, we needed an activity notion (our task description) and a case notion (our case studies). To define the ``timestamp'', we encoded the order in which the tasks occurred in the case studies, starting with 1 for the first task. Based on that, we interpreted the tasks as an event log, where the corresponding case study is the case ID and the encoded order is the timestamp. 
The activity definition was less straightforward: If we considered all dimensions of the task descriptions, we would have hardly found any patterns, due to the large variety in the tasks. Therefore, we first chose just the \textit{task goal} as the activity. This provided a high-level overview over the tasks, outlining how the different goals relate to each other. 
To break this down, we created a second event log, where the tuple (\textit{task goal, data target}) was interpreted as the activity. This showed how the different granularity levels of an analysis occur in a typical conformance checking analysis. 
For both activity notions, we discovered a process model using the process mining tool Disco.\footnote{\url{http://www.fluxicon.com/disco}}
The discovered models are shown in \autoref{fig:task_goal} and \autoref{fig:task_goal_target}.\footnote{\autoref{fig:task_goal_target} only contains the most frequent paths in the model to ensure visual clarity.}

When examining \autoref{fig:task_goal}, we can see how the task goal of the conformance analysis develops over time. It is visible that most conformance analyses aim to \textit{describe} process conformance in the beginning (17 of 33). Alternatively, some analyses (9) start by \textit{exploring} the conformance. The task goal \textit{confirm} only occurs as a start activity and is never used as a follow-up to another task goal, which contradicts how this is typically done in visualization \cite{schulz2013design}. 
After the initial task, we see that the three goals \textit{describe}, \textit{present}, and \textit{explore} often follow. The order of these three is rather flexible, indicated by the loops between them. The last task goal, \textit{explain}, is typically located at the end of the analysis as it is the end activity in the event log in 9 out of 14 occurrences. 


To get a more detailed view on dependencies between tasks, we analyze the task goal in combination with the data target, which corresponds to the granularity level of the analysis (i.e., log, trace, or event). In \autoref{fig:task_goal_target}, we see that the granularity level typically increases over the course of an analysis: It starts on the level of the event log and then moves through the intermediate level of traces to the fine granularity of individual events. This evolving from broad to fine granularity also explains the concurrency of the three goals \textit{describe}, \textit{present}, and \textit{explore}, which we saw in the previous analysis. Concretely, we see that the goals \textit{describe} and \textit{present} are typically applied to an event log first, after which individual traces are analyzed. Further, the goals \textit{describe} and \textit{explore} are often applied on event level right after an application of trace level of the same goal. After that, we see the location of the task \textit{explain} towards the end of the analysis.

\update{R2.3}{
\section{Illustrative Visualizations of Exemplary Tasks}
\label{sec:illustrative}

The overall objective of our task taxonomy is to provide a deeper understanding of the conformance checking domain \cite{munzner2009nested}, with the eventual objective to develop novel high-quality visualizations that are useful in practice and support organizations in leveraging the potentials of conformance checking for their processes. 
Hence, the most important property of our task taxonomy is its ``real-world'' nature, i.e., whether the identified tasks are indeed relevant data analysis questions in practical applications \cite{kerracher2017constructing}.
In this section, we further validate this property by providing ``illustrative concrete examples [of tasks] from multiple domains''~\cite[p.55]{kerracher2017constructing}.
Concretely, we show how exemplary tasks from our taxonomy are visualized in commercial process mining tools. 
These tools were not part of our original task generation, so these exemplary visualizations demonstrate that the tasks that we have identified actually occur in a real-world context. 

\begin{table}[hbt]
\centering
\caption{Exemplary Tasks Illustrated in this Section} 
\label{tab:exemplary}
\begin{tabularx}{0.9\textwidth}{lc}
\toprule
\centering
Task & Frequency \\ 
 & (in Lit.~Review) \\ \midrule
Describe: Derive Process conformance & 14 \\
Present: Present Guideline violations & 10 \\
Explore: Identify Guideline violations & 10 \\
\midrule
Describe: Compare Process conformance & 7 \\
Describe: Summarize	Guideline violations & 4 \\
\midrule
Describe: Summarize	Process conformance	& 2 \\
Describe: Derive Process conformance over time & 1 \\
\bottomrule
\end{tabularx}
\end{table}

As a validation of all 101 tasks is out of scope for this paper, we decided to show visualizations for a selection of tasks that exhibited different levels of frequency in our literature review. 
Specifically, we selected the three most frequent tasks, two tasks with a medium frequency, and two infrequent tasks (see \autoref{tab:exemplary}). 
The illustrative visualizations per task stem from different commercial process mining tools and are briefly explained in the following.

\subsection{Describe: Derive Process Conformance}

As described above, ``Describe: Derive Process Conformance'' relates to the presentation of log-wide fitness values. 
An exemplary visualization from the tool myInvenio \cite{PM20} is shown in \autoref{fig:invenio_fit}.
We match this visualization to this task, as it explicitly presents a derived (i.e., computed) fitness value as a measure for overall conformance between log and model, complemented by minimum and maximum (trace) fitness. 
This visualization relates to control-flow constraints, the data target is a single log. 

\begin{figure}[htb]
    \centering
    \includegraphics[width=.5\textwidth]{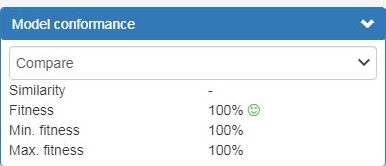}
    \caption{Visualization ``Describe: Derive Process Conformance'' from myInvenio \cite{PM20}}
    \label{fig:invenio_fit}
\end{figure}


\subsection{Explore: Identify Guideline violations}

``Explore: Identify Guideline violations'' provides the user with means to explore the data and independently identify violations.
In the tool myInvenio, shown in \autoref{fig:invenio_graph}, this task is visualized by means of a process model in the form of a directly-follows graph \cite{aalst_2022_overview}. 
Activities that are involved in violations are highlighted in red and the user can further explore these violations by hovering or clicking on them.  
Again, this visualization relates to control-flow constraints, the data target is a single log. 

A second visualization for the same task from the tool UiPath is shown in \autoref{fig:uipath_graph}. 
Again, we see a process model in the form of a directly-follows graph, with individual activities highlighted in color. 
Users are invited to further explore these activities to identify the violations that they belong to. 

\begin{figure}[p]
    \centering
    \includegraphics[width=.5\textwidth]{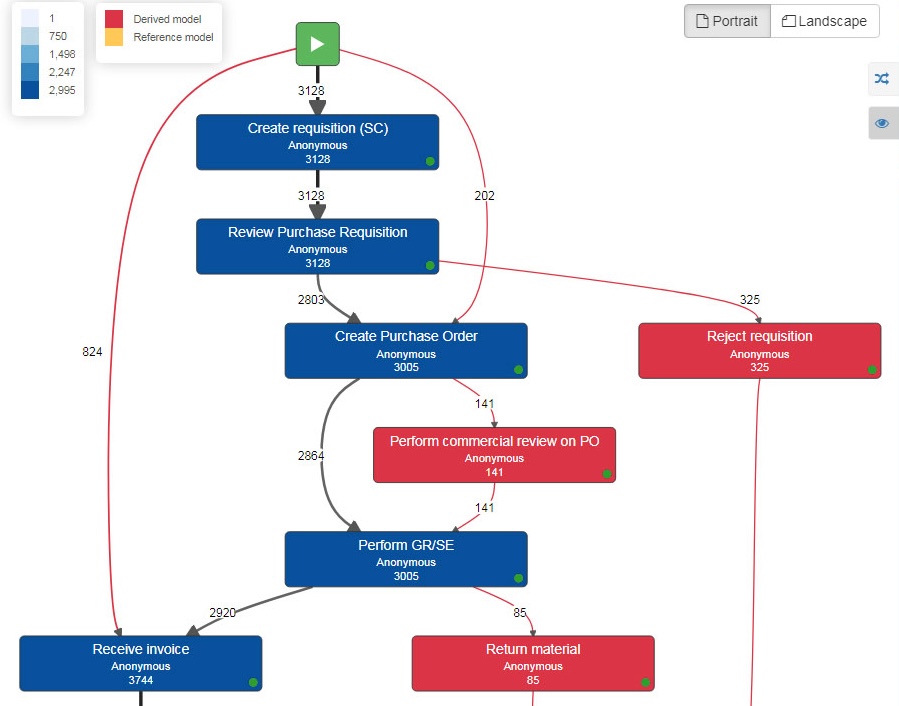}
    \caption{Visualization ``Explore: Identify Guideline violations'' from myInvenio \cite{PM20}}
    \label{fig:invenio_graph}
    \vspace{2em}
    \includegraphics[width=.5\textwidth]{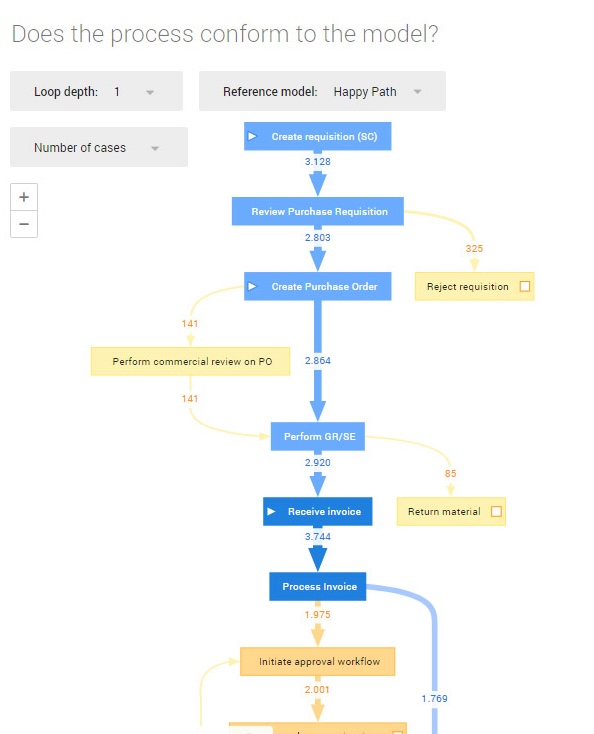}
    \caption{Visualization ``Explore: Identify Guideline violations'' from UiPath \cite{PM20}}
    \label{fig:uipath_graph}
\end{figure}



\subsection{Describe: Compare Process conformance}

``Describe: Compare Process conformance'' builds on ``Describe: Derive Process conformance'' and compares computed fitness values for multiple entities.
In \autoref{fig:aris_conformance_rate_2}, we show an exemplary visualization for this task from the tool ARIS~\cite{PM20}.
This bar chart compares the conformance of multiple event logs from different purchasing organizations within the company, pointing out which overall values are still acceptable (green) and which are not (yellow). 

\begin{figure}[htb]
    \centering
    \includegraphics[width=.4\textwidth]{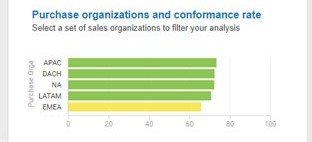}
    \caption{Visualization ``Describe: Compare Process conformance'' from ARIS \cite{PM20}}
    \label{fig:aris_conformance_rate_2}
\end{figure}


\autoref{fig:aris_conformance_rate_3} shows a visualization for the same task, but with different data targets. 
Specifically, this bar chart compares the frequency and the conformance rate of multiple trace variants within one log.
Again, this comparison task relates multiple previously derived conformance values, but compared to the previous one, it works on traces instead of full logs.

\begin{figure}[htb]
    \centering
    \includegraphics[width=.4\textwidth]{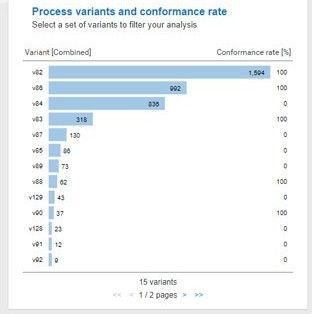}
    \caption{Visualization ``Describe: Compare Process conformance'' from ARIS \cite{PM20}}
    \label{fig:aris_conformance_rate_3}
\end{figure}


\subsection{Describe: Summarize Guideline violations}

``Describe: Summarize Guideline violations'' differs from ``Describe: Derive Process Conformance'' in that it adds up individual guideline violations, not compute a (relative) degree of conformance. 
\autoref{fig:aris_diagnosis_1} shows a corresponding visualization from ARIS \cite{PM20}.
We see the total number of conformance issues in a single log, aggregated into a single number.

\begin{figure}[htb]
    \centering
    \includegraphics[width=.4\textwidth]{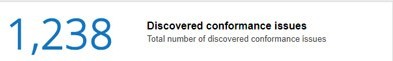}
    \caption{Visualization ``Describe: Summarize Guideline violations'' from ARIS \cite{PM20}}
    \label{fig:aris_diagnosis_1}
\end{figure}


\subsection{Describe: Summarize Process conformance}

``Describe: Summarize Process conformance'' builds on the previous task  ``Describe: Summarize Guideline violations'' and relates the aggregated number of guideline violations to the total number of traces. 
In \autoref{fig:aris_conformance_rate_1}, we see a corresponding visualization from ARIS. This number describes the degree of conformant traces within one log, i.e., the ratio of traces that do not contain a violation. 
This value may differ significantly from the overall degree of conformance as given by a fitness measure, depending on how the violations are distributed among the individual traces. 
A second visualization for the same task from Celonis is shown in \autoref{fig:celonis_overview_2}.
Again, we see the percentage of conformant traces within one log, represented by a percentage value. 

\begin{figure}[htb]
    \centering
    \includegraphics[width=.4\textwidth]{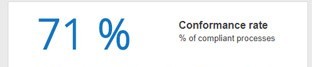}
    \caption{Visualization ``Describe: Summarize Process conformance'' from ARIS \cite{PM20}}
    \label{fig:aris_conformance_rate_1}
\end{figure}


\begin{figure}[htb]
    \centering
    \includegraphics[width=.3\textwidth]{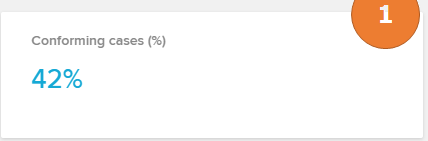}
    \caption{Visualization ``Describe: Summarize Process conformance'' from Celonis \cite{celonisAnalysisConformance}}
    \label{fig:celonis_overview_2}
\end{figure}


\subsection{Describe: Derive Process conformance over time}

Despite occurring only two times in our literature review, ``Describe: Derive Process conformance over time'' is visualized in process mining tools, for example UiPath, as seen in \autoref{fig:UiPath_history}.
This line diagram shows how the absolute amount of non-conforming traces develops over time. 
For comparison, the development of the absolute amount of traces that follow the happy path, i.e., the most frequent trace variant is also shown.
In this diagram, we see that overall, the amount of non-conforming traces is higher, but the development over time differs more gravely than for the happy path cases.

\begin{figure}[htb]
    \centering
    \includegraphics[width=.5\textwidth]{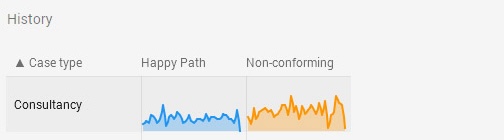}
    \caption{Visualization ``Describe: Derive Process conformance over time'' from UiPath \cite{PM20}}
    \label{fig:UiPath_history}
\end{figure}


\subsection{Present: Present Guideline violations}

Finally, ``Present: Present Guideline violations'' also occurs in commercial tools, as shown by the visualization from myInvenio in \autoref{fig:invenio_devs}.
We see that this visualization lists individual guideline violations on the activity level, naming activities as unexpected and explaining the impact of that violation on the overall process. 
The difference to ``Explore: Identify Guideline violations'' above is that here, the violations are explicated, whereas above, the process model needs to be explored in order to find them.

\begin{figure}[htb]
    \centering
    \includegraphics[width=.5\textwidth]{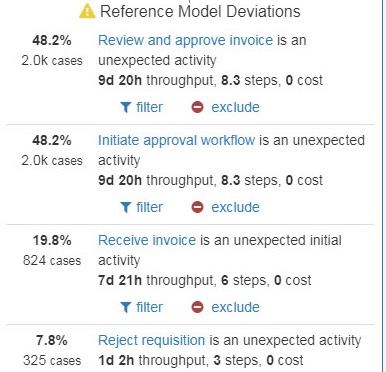}
    \caption{Visualization ``Present: Present Guideline violations'' from myInvenio \cite{PM20}}
    \label{fig:invenio_devs}
\end{figure}


}

\section{Related Work}
\label{sec:related}

In this paper, we bring together visual analytics and process mining by developing a taxonomy for \update{R1.2}{tasks of the conformance checking domain}. In this sense, our work relates to others that are located at the intersection of these two fields. So far, there are no concrete task taxonomies for the process mining domain. However, researchers have explored the potential benefits that the two disciplines present to one another. They have also identified a need for specifying concrete tasks in the process mining domain. 

\paragraph{Process Mining \& Visual Analytics}
Extant work has identified the potential of integrating perspectives from visual analytics into process mining. Van der Aalst et al.~\cite{van_der_aalst_2011_breathing} approach the topic by providing an overview of the combination of the two fields and exploring the use of visual analytics to enhance the understanding of process models. Kriglstein et al.~\cite{kriglstein_2016_visual} categorize process mining techniques based on visual analytics aspects. Gschwandtner~\cite{gschwandtner_2017_visual} discusses challenges and opportunities for process analysis with visual analytics, also with regard to conformance checking. She argues that visual analytics can provide a benefit for many process mining tasks, and recommends an iterative approach for designing visualizations that involves user feedback already at an early stage. Furthermore, visualizations are noted to be highly task-dependent.
\update{R1.2}{This is reflected in the viewpoint put forward by Miksch et al.~\cite{miksch_2024_visual}, who see benefits in the combination of visual analytics and process mining in increasing insights and understanding generated by process mining techniques. In particular, they argue that a taxonomy that combines analytical process mining tasks with concepts from visual analytics could help in the design, research, and evaluation of process visualizations, and would aid in establishing a common understanding of the domain at hand. Notably, they acknowledge that such a taxonomy has not yet been established.}

\paragraph{The Need for Task-Dependent Visualizations}
Echoing the finding of a need for task-dependent visualization, Klinkmüller et al.~\cite{klinkmuller_2019_mining} find that process mining analyses are often visualized with generic visualization techniques not specific to process mining, and that tasks with different information needs are addressed with the same visual representations. The authors further highlight the need for more specific analytical support. In a similar vein, Mendling et al. \cite{mendling2021CognitiveEffectiveness} observe that existing research on process mining ignores the relation between a user's task and the output or visualization of process mining tools. Adding to that, our own previous work \cite{rehse_2022_visualization} analyzes conformance checking visualizations of commercial and academic process mining tools, and finds that tools rely on generic visualization idioms and that there is a lack of research on information needs and user preferences.

\paragraph{Event Sequence Analysis}
More broadly, process mining relates to event sequence analysis. Guo et al. \cite{guo2022SurveyEventSequence} propose a taxonomy for visual analytics techniques of event sequence data, in which they extract five high-level analytical tasks common to these techniques: \textit{summarization} (aiming to uncover patterns and progressions), \textit{event prediction and recommendation} (using historical data to predict the future and support decision-making), \textit{anomaly detection} (identifying deviations from the usual progression of events), \textit{comparison} (identifying similarities and differences of event sequences), and \textit{causality analyses} (exploring and understanding causal relations). For these tasks, according to the study, various techniques and means exist. Building on this taxonomy, Yeshchenko and Mendling \cite{yeshchenko_2024_survey} provide a survey of and categorization framework for visualization techniques for event sequences in the information visualization and process mining fields. They find that process mining research focuses on generating formal process models, and information visualization on interaction with visualizations in specific application domains. They also highlight potential synergies for both fields.

\paragraph{Visualizations for Conformance Checking}
In addition to these abstract analyses, some studies also propose concrete approaches and visualizations in the area of process mining and conformance checking.
An approach for identifying bottlenecks in business processes through the analysis of event logs and visualization of the results is proposed by Kaouni et al. \cite{kaouni_2021_visual}. Their study aims to provide explainable and interpretable results for business processes based on visual analytics, without overwhelming users with complex process models.
Gall and Rinderle-Ma \cite{gall_2021_3Dconf} develop visualizations for instance/process-spanning compliance constraints, and find that expert users prefer color-based representations for gaining an overview and text-based representations for conducting in-depth analyses.
Furthermore, in our own previous work \cite{klessascheck2024StructuralModel}, we design a structural model of conformance checking visualizations for deriving descriptive models for concrete analyses. The structural model is evaluated in a user study with concrete conformance checking tasks, which shows that the resulting visualizations are appropriate and accessible. Notably, generic visualizations such as pie charts were perceived as not appropriate by users.
Moreover, Knoblich et al. \cite{knoblich2024VisualEncodings} review visualizations of six commercial process mining tools. Using an extended taxonomy for visualizations of event sequences, they conduct a user study to assess the effectiveness of the different tools and their visualizations of process models. They find that there is room for improvement across all investigated tools, especially in the comprehensible use of color, the legibility of annotations, and the understandability of marks and channels (i.e., geometric elements and their appearance that are used to encode data \cite{munzner2014visualization}). 
Finally, Brinkman et al. \cite{brikman2024interpretability} conduct a user study on visualizations of discovered process models along several user tasks. Several issues that hamper the interpretability of discovered process models are identified, and requirements of process model visualizations to encounter these challenges are derived.

\bigskip
\noindent
To summarize, existing studies agree that integrating visual analytics aspects and a focus on user concerns into process mining and conformance is beneficial because it supports end users in their analysis tasks. However, it also becomes clear that the body of related work has not yet found a way to formulate conformance checking tasks in a sufficiently explicit manner.

\section{Discussion \& Conclusion}
\label{sec:conclusion}

In this paper, we introduce a task taxonomy for conformance checking, which classifies tasks along six dimensions: task goal, task means, data characteristics, constraint type, data target, and data cardinality. Based on a literature review of 33 conformance checking case studies, we identified 102 tasks, of which ``Describe: Derive Process conformance'', ``Present: Present Guideline violations'', and ``Explore: Identify Guideline violations'' occurred most frequently. By analyzing dependencies between the tasks by means of process discovery, we found that conformance checking analyses are typically conducted in an exploratory rather than a confirmatory fashion and that during an analysis, the level of granularity increases, from a log-level overview to an event-level analysis of individual violation patterns.

\secondupdate{R2}{

\subsection{Contributions \& Future Work }

This work was motivated by the observation that, despite the wide variety of existing visualizations for conformance checking, their efficacy often remains unclear. 
It is difficult for researchers to evaluate these visualizations because we often lack a clear understanding of the analytical purpose they are intended to serve. 
Our task taxonomy addresses this issue by providing a structured account of the tasks that users aim to perform during conformance checking.
By characterizing them along the six selected dimensions, we provide a detailed characterization of the individual tasks, which indicates which information is required and not required to fulfill them. 

With this taxonomy in place, visualizations can now be evaluated more systematically: rather than assessing them only in general terms, we can employ empirical methods to evaluate to what extent they support specific tasks within the taxonomy. 
For example, we can examine whether a given visualization enables users to derive the process conformance, identify guideline violations, or compare conformance across different process variants. 
This task-based perspective allows for a more meaningful evaluation of existing visualizations and provides a reference point for identifying potential gaps in current tool support. 
Specifically, the taxonomy can 
\begin{inparaenum}[(1)]
    \item support the evaluation of existing visualizations by enabling a systematic comparison of supported tasks, 
    \item inform the design of new visualizations by highlighting which tasks should be supported and how analysis workflows can be aligned with user needs, and 
    \item guide the evaluation of new and existing visualizations by providing a structured basis for empirical studies in which users perform defined tasks.
\end{inparaenum}

As an example, consider the task of identifying guideline violations (``Explore: Identify Guideline violations''), which occurred 10 times in our literature review and is present in many process mining tools, although with different supporting visualizations. 
In an empirical study for this task, participants could be asked to use one of those visualization to analyze an event log, with the goal to detect and report violations of predefined process guidelines. 
The efficacy of the violation could be measured in terms of accuracy in identifying violations (effectiveness) and required time (efficiency),  alongside qualitative feedback on usability. 
By structuring the study around the task taxonomy, researchers ensure the evaluation focuses on the visualization’s efficacy in supporting this specific analytical purpose, which provides them with a clear roadmap for future evaluations. 

In addition, the taxonomy can guide the design of new visualizations by linking design decisions to concrete user needs. It offers a way to reason about which conveyed information and which visual idioms are useful and appropriate for which analytical tasks, thereby helping to move beyond ad-hoc design choices. In this sense, our contribution provides a conceptual foundation for both evaluating and designing conformance checking visualizations in a more purposeful and user-centered manner.

\subsection{Limitations \& Threats to Validity}

Our work is subject to multiple limitations. 
First, our collection of tasks is based on a review of case studies, which are a specific type of scientific paper. To ensure broad coverage of potentially relevant literature, we built on the results of a previous literature review \cite{emamjome_2019_case_study}, which we extended with another literature review that also covered papers published after 2018. Still, we cannot expect our list of papers to be comprehensive. This is already difficult to achieve for regular papers \cite{vom2015standing}, but it is even more complicated for case studies, which are often published as evaluations to novel technical contributions instead of as stand-alone papers. This means that we likely missed case studies that were included in other research papers or visualization tasks that occurred in papers that were not published as case studies. 

Second, with respect to the taxonomy development, our research method relied on our subjective judgment and thus might potentially be biased. 
We tried to mitigate this by designing the method according to Kerracher et al.~\cite{kerracher_2015_task} and Nickerson et al.\cite{nickersonMethodTaxonomyDevelopment2013} and documenting each step in \autoref{sec:method}.
In our method, we further tried to mitigate the influence of subjective bias by having each task coded by two authors individually. Potential disagreements were discussed in the full author group until a consensus was reached. Nevertheless, some risk of subjective bias remains as a potential limitation.

Third, our taxonomy is designed for tasks to be independent from a concrete conformance checking technique, meaning that in principle, any technique can be used to conduct any task. 
As pointed out above, we made this decision purposefully, to ensure the broad applicability of our taxonomy.
However, individual tasks that we found in the case studies may still have been influenced by the underlying technique. 
For example, alignments typically identify violations on the level of individual activities, where as rule checking may detect violations consisting of multiple activities. 
This will also become relevant when developing visualizations for the respective tasks.

From these limitations, we can conclude that our list of identified tasks is most likely not complete. Judging from our domain knowledge and the underlying design space of visualization tasks \cite{schulz2013design,tominski2020interactive}, we assume the four dimensions \textit{task goal}, \textit{constraint type}, \textit{data target}, and \textit{data cardinality} to be complete with regard to their characteristics, but the taxonomy allows for numerous combinations of these values, which were not observed in our task list. 
In addition, the extensive list of characteristics that we found for the \textit{task means} and \textit{data characteristics} dimensions makes it rather likely that there are other characteristics, which we did not cover in our current task list. 
Essentially, we consider our list to be a first observation of tasks that can occur in a conformance checking analysis. 
It could be extended by exploring conformance checking visualization tasks through other methods, such as interviews with domain experts~\cite{kerracher_2015_task}.

Besides completeness, another relevant aspect is the validity of our identified tasks.
As pointed out above, this mainly concerns ensuring that our tasks are of ``real-world'' nature, meaning that they actually describe relevant data analysis questions.
In this paper, we tried to ensure this real-world nature by relying on case studies that applied conformance checking on real-world event data in a practical setting.
This implies that all identified tasks have been relevant in at least one practical application. 
We further provided evidence for the real-world nature of tasks by providing illustrative concrete examples of task visualizations from commercial process mining tools in \autoref{sec:illustrative}. 
The fact that tasks with differing frequencies in our case studies were included in those tools shows that they are relevant to the customers of those tools, further underlining their practical relevance.
However, as we were only able to provide these examples for a selected number of cases, further validation of our taxonomy is necessary and desirable in future work. 

\subsection{Conclusion}

Despite these limitations, our work can make an important contribution to both process mining and visual analytics. 
By applying a research method from the visualization community to the process mining field, we develop a deeper understanding of conformance checking, from which both fields can benefit. 
We envision our taxonomy to serve as a foundation for future research in the intersection of process mining and visualization, including, but not limited to, the above-mentioned evaluation of existing visualizations and the design of new ones.

In addition, the deeper understanding of conformance checking can also provide added value to the domain itself.
For example, the taxonomy structures the questions that conformance checking techniques can answer, which facilitates systematic comparisons.
It also highlights underexplored areas in conformance checking research, giving hints for novel and potentially interesting research ideas.
This way, our work can foster both the practical adoption of conformance checking techniques as well as increased collaboration between researchers from the two fields.

}

\paragraph*{Data Availability}
All our research data (list of case studies, derived list of tasks, task descriptions, task analyses) is available at \url{https://figshare.com/s/2c2c505ac602c17d89fe}. If the paper is accepted, we plan to convert this link into a permanent DOI.

\paragraph*{Project Funding}
Our work was partially funded by DFG grant 465904964 ``Conformance Checking With Regulations''. It has benefited substantially from Dagstuhl Seminar 223271 ``Human in the (Process) Mines''.

 \bibliographystyle{elsarticle-num} 
 \bibliography{bibliography}





\end{document}